\documentclass[aps,a4paper,10pt,twocolumn,nofootinbib,floatfix]{revtex4} 
\usepackage[T1]{fontenc}
\usepackage{mathptmx}
\usepackage{datetime}
\usepackage{amsmath}
\usepackage{amsfonts}
\usepackage{amssymb}
\usepackage{mathrsfs}
\usepackage[mathscr]{euscript}
\usepackage[dvips]{graphicx}
\usepackage{fancyhdr}
\usepackage{hyperref} 
\usepackage{makeidx}
\usepackage{color}

\def\overstrike#1#2{{\setbox0\hbox{$#2$}\hbox to \wd0{\hss
    $#1$\hss}\kern-\wd0\box0}}

\newcommand{\convol}{\star}
\newcommand{\grad}{\nabla}
\newcommand{\cross}{\times}

\renewcommand{\Vec}{\textbf}

\newcommand{\pLightspeed}{c}

\newcommand{\pPermittivity}{\epsilon}   
\newcommand{\pPermittivityVac}{\epsilon_0} 

\newcommand{\pPermeability}{\mu}         
\newcommand{\pPermeabilityVac}{\mu_0}      

\newcommand{\pXemXelectric}{E}                      
\newcommand{\pXemXelectricv}{\Vec{\pXemXelectric}}  
\newcommand{\pXemXdisplacement}{D}                         
\newcommand{\pXemXdisplacementv}{\Vec{\pXemXdisplacement}} 
\newcommand{\pXemXpolarization}{P}                         
\newcommand{\pXemXpolarizationv}{\Vec{\pXemXpolarization}} 

\newcommand{\pXemXmagnetic}{B}                           
\newcommand{\pXemXmagneticv}{\Vec{\pXemXmagnetic}}       
\newcommand{\pXemXmagstrength}{H}                        
\newcommand{\pXemXmagstrengthv}{\Vec{\pXemXmagstrength}} 
\newcommand{\pXemXmagnetization}{M}                          
\newcommand{\pXemXmagnetizationv}{\Vec{\pXemXmagnetization}} 

\newcommand{\pCurrent}{J}                  
\newcommand{\pCurrentv}{\Vec{\pCurrent}}   

\newcommand{\pEfield}{\pXemXelectric}
\newcommand{\pDfield}{\pXemXdisplacement}

\newcommand{\pBfield}{\pXemXmagnetic}

\newcommand{\pEfieldv}{\pXemXelectricv}
\newcommand{\pDfieldv}{\pXemXdisplacementv}
\newcommand{\pPfieldv}{\pXemXpolarizationv}
\newcommand{\pBfieldv}{\pXemXmagneticv}
\newcommand{\pHfieldv}{\pXemXmagstrengthv}
\newcommand{\pMfieldv}{\pXemXmagnetizationv}

\newcommand{\pKcurrentv}{\Vec{K}} 

\newdateformat{yymmdddate}{\THEYEAR/\twodigit{\THEMONTH}/\twodigit{\THEDAY}}

\numberwithin{equation}{section}

\renewcommand{\theequation}{\arabic{section}.\arabic{equation}}

\newcommand{\XDOI}[1]{\href{http://dx.doi.org/#1}{doi:#1}}
\newcommand{\XARXIV}[1]{\href{http://arxiv.org/abs/#1}{arXiv:#1}}

\begin{document}

\title{Uni-directional optical pulses and temporal propagation: \\
 with consideration of spatial and temporal dispersion}

\author{P. Kinsler}
\homepage[]{https://orcid.org/0000-0001-5744-8146}
\email[\hphantom{.}~]{Dr.Paul.Kinsler@physics.org}
\affiliation{
  Physics Department,
  Lancaster University,
  Lancaster LA1 4YB,
  United Kingdom.}

\affiliation{
  Blackett Laboratory, Imperial College London,
  Prince Consort Road,
  London SW7 2AZ,
  United Kingdom.}

\begin{abstract}

I derive a temporally propagated uni-directional optical pulse equation
 valid in the few cycle limit.
Temporal propagation is advantageous
 because it naturally preserves causality, 
 unlike the competing spatially propagated models.
{The exact coupled bi-directional equations
 that this approach generates}
 can be efficiently approximated down to a uni-directional form 
 in cases where an optical pulse changes little over
 one optical cycle.
{They also permit a direct term-to-term comparison
 of the exact bi-directional theory with 
 its corresponding} approximate uni-directional theory.
Notably, 
 temporal propagation
 handles dispersion in a different way, 
 and this difference serves to highlight existing approximations
 inherent in spatially propagated treatments of dispersion.
Accordingly, 
 I emphasise the need for future work in clarifying the limitations
 of the dispersion conversion required by these types of approaches; 
 since the only alternative in the few cycle limit 
 may be to resort to 
 the much more computationally intensive
 full Maxwell equation solvers. \\

\end{abstract}

\lhead{\includegraphics[height=5mm,angle=0]{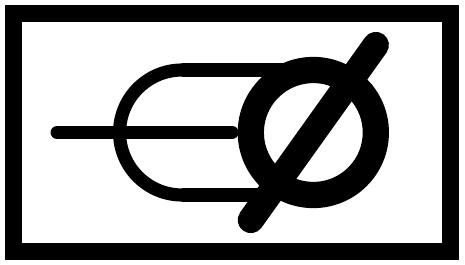}~~D2OWE}
\chead{} 
\rhead{
\href{mailto:Dr.Paul.Kinsler@physics.org}{Dr.Paul.Kinsler@physics.org}\\
\href{http://www.kinsler.org/physics/}{http://www.kinsler.org/physics/}
}

\date{\today}
\maketitle
\thispagestyle{fancy}

\section{Introduction}
\label{S-Introduction}

The importance of having simple and robust methods 
 for the propagation of optical pulses
 has attracted increasing attention in recent years.
This is due to the multitude of applications \cite{Brabec-K-2000rmp}
 {in which ever shorter pulses
 are used to act like a strobe-lamp that
 takes snapshots of ultrafast processes
 \cite{Corkum-1993prl,Schafer-YDK-1993prl},
 or sub-wavelength electric field profiles are created
 \cite{Fuji-RGAUYTHK-2005njp,Radnor-CKN-2008pra,Solanpaa-BSCRR-2014pra}
 to achieve detailed control over atomic or molecular responses.
Alternatively, 
 strong nonlinearity can be used to construct pulses that 
 are both wide-band and temporally extended, 
 such as (white light) supercontinnua 
 \cite{Alfano-S-1970prl,Alfano-S-1970prl-2,Dudley-GC-2006rmp}.
Such nonlinearities can even be used 
 to generate sub-structure that is instead} temporally confined,
 as in optical rogue waves \cite{Solli-RKJ-2007n}; 
 or even the temporally and spatially localized 
 filamentation processes \cite{Braun-KLDSM-1995ol,Chin-PBM-1999jjapl,Milchberg-CCJPRVWZ-2014pp}.

{This progress towards achieving ever shorter pulse durations,
 with their associated larger spectral bandwidths, 
 and higher pulse intensities,
 has been pushing traditional
 pulse propagation models to their limits, 
 or breaking them.
If we want to avoid the computational expense of always resorting to
 high resolution Maxwell's equations solvers
 coupled to detailed material response models, 
 we need to be confident that our 
 simpler, 
 less demanding approaches still work.
In particular 
 we need to have a clear idea of the physics that may have been removed,
 and what the side-effects of those approximations are.
In this paper
 I use a directional approach, 
 whose relatively simple and straightforward derivation
 allows easy comparison of both the approximate and the exact
 propagation equations, 
 whilst still resulting in the analytical and numerical
 convenience of a first-order wave equation.}

{However, 
 unlike {perhaps the most common approach in nonlinear optics}, 
 I consider propagation in time rather than along some chosen
 spatial axis  \cite{Kinsler-2010pra-fchhg,Kinsler-2012arxiv-fbrad}.
The most significant distinctions between temporal and spatial propagation
 approaches are summarized
 on figs. \ref{fig-reflect-t} and \ref{fig-reflect-z}
 respectively.
Notice that the initial conditions (starting states)
 are completely different, 
 and that only the temporal propagation model is going to provide
 causal solutions in a straightforward way.
The temporally propagated evolution equations for the spatial wave profiles
 do not give us access
 to the frequency spectrum \cite{Kinsler-2014arXiv-negfreq}, 
 but instead we have the wavevector spectrum, 
 as well as a causally appropriate time evolution 
 \cite{Kinsler-2011ejp}.
This requires an alternate approach to the temporal response 
 of the propagation medium, 
 which highlights the existing
 and often somewhat poorly characterised approximations
 inherent in spatially propagated treatments of dispersion.
Of course, 
 full finite element and/or FDTD \cite{Yee-1966tap} pulse propagation
 can also be used, 
 but here my aim is to simplify the time propagated approach
 in a directional approach in line with common spatially propagated methods.}

\def\xtcol{}

\begin{figure}
\includegraphics[angle=-0,width=0.82\columnwidth]{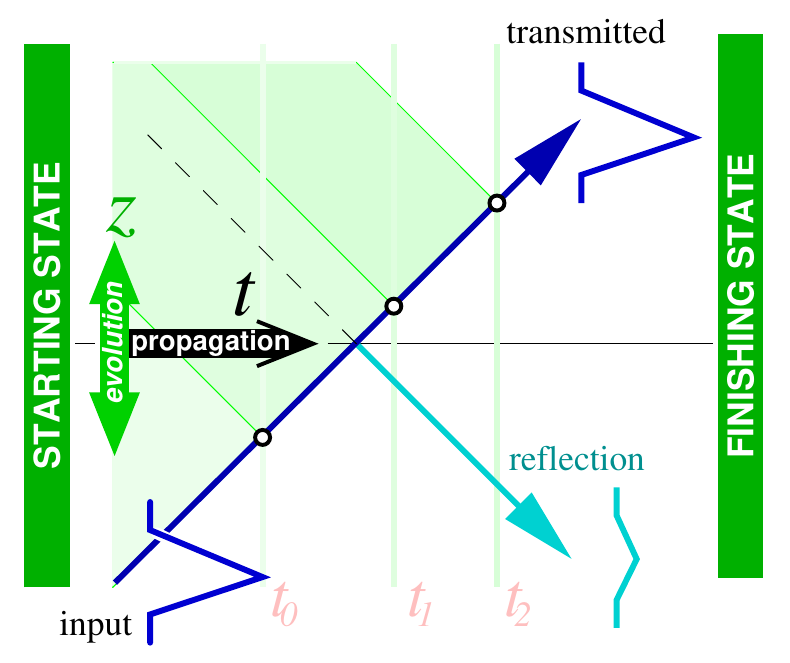}
\caption{
 Temporal propagation of waves, 
  where disturbances (or pulses)
  evolve either forward or backward in space.
 At any point in the propagation, 
  we know the spatial behaviour of our wave field at all points in space, 
  as indicated by the pale{\xtcol} vertical lines.
 The pale{\xtcol} shaded triangles indicate the past light cones
  (i.e. the causal past)
  of a wave element (black circles)
  at selected times  
  along the path of the disturbance.
 In a temporally propagated numerical simulation
  which has a maximum wave speed,
  the causal past matches the computational past.
 A notional interface has been added to the diagram
  to show how a reflection would behave.
Figure used with permission from \cite{Kinsler-2014arXiv-negfreq}.
}
\label{fig-reflect-t}
\end{figure}

\def\xzcol{}

\begin{figure}
\includegraphics[angle=-0,width=0.82\columnwidth]{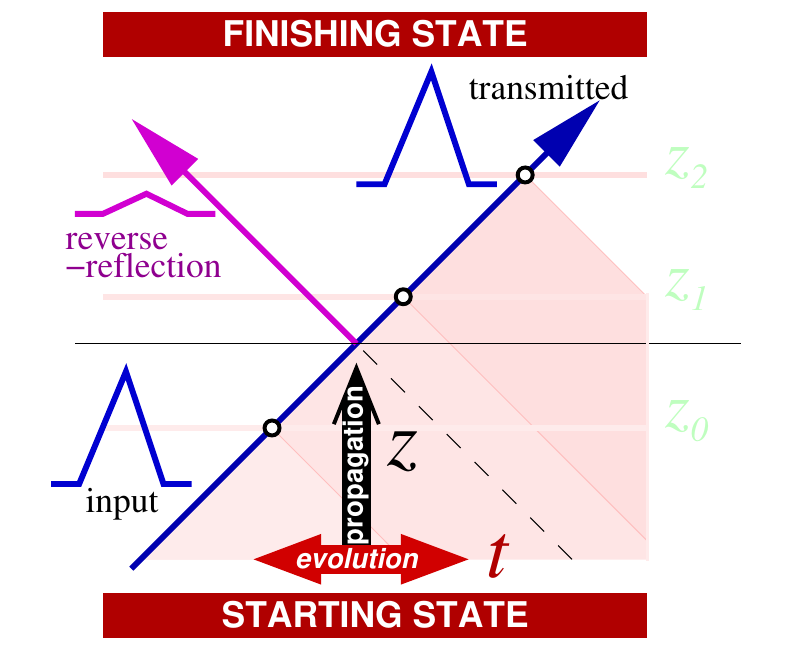}
\caption{
 Spatial propagation of waves, 
  where disturbances (or pulses)
  evolve either forward or backward in time.
 At any point in the propagation, 
  we know the full time behaviour of our wave field --
  both history and future, 
  as indicated by the pale{\xzcol} horizontal lines..
 The light{\xzcol} shaded triangles indicates the computational past 
  of a wave element at particular points (black circles) 
  along the path of the disturbance.
 Note that unlike the temporally propagated case
  shown in fig. \ref{fig-reflect-t}, 
  the computational past of a spatially propagated system is not the same 
  as the causal past.
 A notional interface has been added to the diagram
  to show how reflections behave --
  i.e. in an unexpected way \cite{Kinsler-2012arXiv-fbacou}.
 This is because a reflection should
  have been put in the initial conditions, 
  but was not due to an (assumed) lack of knowledge of that future behaviour.
Figure used with permission from \cite{Kinsler-2014arXiv-negfreq}.
}
\label{fig-reflect-z}
\end{figure}


In section \ref{S-secondorder}
 I derive a customized second order wave equation 
 from Maxwell's equations, 
 and reorganize it to define the material properties appropriately 
 and set up the factorization stage.
{In section \ref{S-factorisation}, 
 I use a factorization method \cite{Kinsler-2010pra-fchhg}
 that allows us to construct an explicitly bi-directional model, 
 and which is then approximated to the popular uni-directional limit 
 in section \ref{S-firstorder}.
Section \ref{S-modifications} remarks on commonly used modifications
 that can be applied to the equations 
 given in sections \ref{S-factorisation} and \ref{S-firstorder},
 typically in order to clarify their properties,
 simplify them,
 or compare them to existing models.}
Having derived this main result -- 
 the propagation equations and their various specializations --
 I turn in {section} \ref{S-dispersion}
 to the consequences of temporal propagation
 and the perspective it provides on the handling of 
 mixed spatial and temporal dispersive effects.
This is then followed in section \ref{S-nls} by a specific comparison between 
 the ordinary spatially propagated nonlinear Schr\"odinger (NLS) equation --
 perhaps the most widely investigated equation in nonlinear optics -- 
 and its temporally propagated counterpart.
Finally, 
 in section \ref{S-conclude}, 
 I present my conclusions.

%
\section{Second order wave equation}
\label{S-secondorder}

{My starting point is 
 the standard macroscopic Maxwell's equations, 
 where I aim for a solution based on 
 a uniform and source free dielectric medium, 
 but still intending to allow 
 material properties that are as general as possible.
A typical approach to this is to construct
 a second order wave equation for the electric field $\pEfieldv$, 
 as results from the substitution of the}
 $\grad \cross {\pHfieldv} = \partial_t {\pDfieldv} + {\pCurrentv}$
 Maxwell's equation into 
 $\grad \cross {\pEfieldv} =  -\partial_t {\pBfieldv}$ 
 (see e.g. \cite{Agrawal-NFO}).
Here, 
 however, 
 I want to follow the displacement field ${\pDfieldv}$ (not ${\pEfieldv}$)
 since it naturally occurs in conjuction with a time derivative --
 likewise, 
 the other important field is ${\pBfieldv}$ rather than ${\pHfieldv}$.
{Thus we rearrange the curl Maxwell's equations 
 to emphasize both their causal properties \cite{Kinsler-2011ejp}
 and their time derivatives,
 as}
~
\begin{align}
  {\pPermeabilityVac}
  \partial_t
    {\pDfieldv}
&=
 {\pPermeabilityVac}
  \grad
 \cross
  {\pHfieldv}
 -
  {\pPermeabilityVac}
  {\pCurrentv}
\quad =
  \grad
 \cross
  {\pBfieldv}
 -
  {\pPermeabilityVac}
  \grad
  \cross
  {\pMfieldv}
 -
  {\pPermeabilityVac}
  {\pCurrentv}
,
\label{eqn-MEalt-curlD}
\\
  {\pPermittivityVac}
  \partial_t
    {\pBfieldv}
&=
 -
  {\pPermittivityVac}
  \grad
 \cross
  {\pEfieldv}
 -
  {\pPermittivityVac}
  {\pKcurrentv}
~~=
 -
  \grad
 \cross
  {\pDfieldv}
 +
  \grad
  \cross
  {\pPfieldv}
 -
  {\pPermittivityVac}
  {\pKcurrentv}
  .
\label{eqn-MEalt-curlB}
\end{align}
Note that for both of these equations,
 we will also (as usual)
 want to specify the material response, 
 whether dielectric or magnetic, 
 in terms of a dielectric polarization ${\pPfieldv}={\pDfieldv}-{\pPermittivityVac} {\pEfieldv}$ 
 and a magnetization ${\pMfieldv}={\pBfieldv}-{\pPermeabilityVac} {\pHfieldv}$.
Either of these equations given above might be either in 
 the spatial domain (with argument $\Vec{r}=(r_x,r_y,r_z)$), 
 or the wavevector domain (with argument $\Vec{k}=(k_x,k_y,k_z)$).

Now we define the total magnetization ${\pMfieldv}$
 in terms of a component ${\pMfieldv}_L$ linearly dependent
 on ${\pBfieldv}$, 
 and another component ${\pMfieldv}_{\textup{\pBfield}}$  containing the rest.
As a result, 
 the excess polarization
 ${\pMfieldv}_{\pPermeability} \equiv {\pMfieldv}_{\textup{\pBfield}}$, 
 which will typically include any nonlinearity, 
 is also a function of ${\pBfieldv}$ and not ${\pHfieldv}$.
In traditional ${\pHfieldv}$-based
 spatially-propagated approaches, 
 the temporal response (``dispersion'') can be incorporated naturally, 
 here it cannot.
We have that 
~
\begin{align}
  {\pMfieldv}(\Vec{r})
&=
  {\pMfieldv}_L(\Vec{r}) + {\pMfieldv}_{\textup{\pBfield}}(\Vec{r}) 
&=
  \alpha_{\pPermeability}(\Vec{r}) {\pBfieldv}(\Vec{r}) + {\pMfieldv}_{\textup{\pBfield}}(\Vec{r})
,
\label{eqn-D2owe-Mr}
\end{align}
 which in the spatial frequency (wavevector $\Vec{k}$) domain is 
~
\begin{align}
  {\pMfieldv}(\Vec{k})
&=
  {\pMfieldv}_L(\Vec{k}) + {\pMfieldv}_{\textup{\pBfield}}(\Vec{k})
&=
  \alpha_{\pPermeability}(\Vec{k}) \convol {\pBfieldv}(\Vec{k}) + {\pMfieldv}_{\textup{\pBfield}}(\Vec{k})
\label{eqn-D2owe-Mk}
,
\end{align}
 since 
 after a Fourier transform from $\Vec{r}$ into $\Vec{k}$, 
 the products become spatial convolutions (``$\convol$''); 
 which emphasises that anything wavevector dependent 
 is necessarily a non-local concept.
Note that 
 this is the converse of the usual situation where
 a temporal response is described using time-domain convolutions,
 and these become products in the frequency domain.

Now we define the total polarization ${\pPfieldv}$
 in terms of a component ${\pPfieldv}_L$ linearly dependent
 on 
 ${\pDfieldv}$, 
 and another component ${\pPfieldv}_{\textup{\pDfield}}$  containing the rest.
As a result, 
 the excess polarization
 ${\pPfieldv}_{\pPermittivity} \equiv {\pPfieldv}_{\textup{\pDfield}}$, 
 which will typically include any nonlinearity, 
 is also a function of ${\pDfieldv}$.
We have that 
~
\begin{align}
  {\pPfieldv}(\Vec{r})
&=
  {\pPfieldv}_L(\Vec{r}) + {\pPfieldv}_{\textup{\pDfield}}(\Vec{r}) 
&=
  \alpha_{\pPermittivity}(\Vec{r}) {\pDfieldv}(\Vec{r})
 +
  {\pPfieldv}_{\textup{\pDfield}}(\Vec{r})
,
\label{eqn-D2owe-Pr}
\end{align}
 which in the spatial frequency (wavevector $\Vec{k}$) domain is 
~
\begin{align}
  {\pPfieldv}(\Vec{k})
&=
  \alpha_{\pPermittivity}(\Vec{k}) \convol {\pDfieldv}(\Vec{k})
 +
  {\pPfieldv}_{\textup{\pDfield}}(\Vec{k})
.
\label{eqn-D2owe-Pk}
\end{align}

With these definitions for $\partial_t {\pDfieldv}$, 
 $\partial_t {\pBfieldv}$, 
 and for ${\pMfieldv}$ and ${\pPfieldv}$ set up, 
 we can now proceed with the derivation of a second order wave equation
 for ${\pDfieldv}$ --
~
\begin{align}
  {\pPermeabilityVac}
  \partial_t
    {\pDfieldv}
&=
  \grad
 \cross
  {\pBfieldv}
 -
  {\pPermeabilityVac}
  \grad
  \cross
  {\pMfieldv}_L
 -
  {\pPermeabilityVac}
  \grad
  \cross
  {\pMfieldv}_{\textup{\pBfield}}
 -
  {\pPermeabilityVac}
  {\pCurrentv}
\\
&=
  \grad
 \cross
  {\pBfieldv}
 -
  \grad
  \cross
  \alpha_{\pPermeability} {\pBfieldv}
 -
  {\pPermeabilityVac}
  \grad
  \cross
  {\pMfieldv}_{\textup{\pBfield}}
 -
  {\pPermeabilityVac}
  {\pCurrentv}
\\
&=
  \grad
 \cross
  {\pBfieldv}
 -
  \alpha_{\pPermeability} 
  \grad
  \cross
  {\pBfieldv}
 -
  \left( \grad \alpha_{\pPermeability} \right)
  \cross
    {\pBfieldv}
 -
  {\pPermeabilityVac}
  \grad
  \cross
  {\pMfieldv}_{\textup{\pBfield}}
 -
  {\pPermeabilityVac}
  {\pCurrentv}
\\
&=
  \grad
 \cross
  \left[
    1
   -
    \alpha_{\pPermeability} 
  \right]
  {\pBfieldv}
 -
  {\pPermeabilityVac}
  \grad
  \cross
  {\pMfieldv}_{\textup{\pBfield}}
 -
  {\pPermeabilityVac}
  {\pCurrentv}
.
\end{align}
Now take the time derivative of this equation, 
 use the fact that $\alpha_{\pPermeability}$ is independent of time,
 define ${a}_{\pPermeability} = 1 - \alpha_{\pPermeability}$
 and ${a}_{\pPermittivity}=1-\alpha_{\pPermittivity}$; 
 and then substitute for $\partial_t {\pBfieldv}$, 
 so that
~
\begin{align}
  {\pPermittivityVac}
  {\pPermeabilityVac}
  \partial_t^2
    {\pDfieldv}
&=
  {\pPermittivityVac}
  \grad
 \cross
  {a}_{\pPermeability} 
  \partial_t {\pBfieldv}
 -
  {\pPermittivityVac}
  {\pPermeabilityVac}
  \grad
  \cross
  \partial_t {\pMfieldv}_{\textup{\pBfield}}
 -
  {\pPermittivityVac}
  {\pPermeabilityVac}
  \partial_t {\pCurrentv}
\\
  c^{-2}
  \partial_t^2
    {\pDfieldv}
&=
 -
  \grad
 \cross
  {a}_{\pPermeability}
  \left[
    \grad
   \cross
    {\pDfieldv}
   -
    \grad
    \cross
    {\pPfieldv}
   +
    {\pPermittivityVac}
    {\pKcurrentv}
  \right]
\nonumber
\\
&\qquad\qquad
 -
  c^{-2}
  \grad
  \cross
  \partial_t {\pMfieldv}_{\textup{\pBfield}}
 -
  c^{-2}
  \partial_t {\pCurrentv}
\\
  \partial_t^2
    {\pDfieldv}
&=
 -
  {\pLightspeed}^2
  \grad
 \cross
  {a}_{\pPermeability}
  \left[
    \grad
   \cross
    {\pDfieldv}
   -
    \grad
    \cross
    \alpha_{\pPermittivity}
    {\pDfieldv}
   -
    \grad
    \cross
    {\pPfieldv}_{\textup{\pDfield}}
   +
    {\pPermittivityVac}
    {\pKcurrentv}
  \right]
\nonumber
\\
&\qquad\qquad
 -
  \grad
  \cross
  \partial_t {\pMfieldv}_{\textup{\pBfield}}
 -
  \partial_t {\pCurrentv}
\\
&=
 -
  {\pLightspeed}^2
  \grad
 \cross
  {a}_{\pPermeability}
  \left[
    \grad
   \cross
    \left( 1 - \alpha_{\pPermittivity} \right)
    {\pDfieldv}
   -
    \grad
    \cross
    {\pPfieldv}_{\textup{\pDfield}}
   +
    {\pPermittivityVac}
    {\pKcurrentv}
  \right]
\nonumber
\\
&\qquad\qquad
 -
  \grad
  \cross
  \partial_t {\pMfieldv}_{\textup{\pBfield}}
 -
  \partial_t {\pCurrentv}
\\
&=
 -
  {\pLightspeed}^2
  \grad
 \cross
  {a}_{\pPermeability}
    \grad
   \cross
    {a}_{\pPermittivity}
    {\pDfieldv}
 +
  {\pLightspeed}^2
  \grad
 \cross
  {a}_{\pPermeability}
    \grad
    \cross
    {\pPfieldv}_{\textup{\pDfield}}
\nonumber
\\
&\qquad\qquad
 -
  {\pLightspeed}^2
  \grad
 \cross
  {a}_{\pPermeability}
    {\pPermittivityVac}
    {\pKcurrentv}
 -
  \grad
  \cross
  \partial_t {\pMfieldv}_{\textup{\pBfield}}
 -
  \partial_t {\pCurrentv}
\label{eqn-D2owe-CX}
\end{align}

Remember that 
 in this picture, 
 all field properties are specified as functions of our chosen
 primary field ${\pDfieldv}$; 
 notably, 
 the magnetic field ${\pBfieldv}$ which is usually
 the main argument for ${\pMfieldv}_{\textup{\pBfield}}$
 must be determined from the calculated ${\pDfieldv}$
 using the relevant Maxwell's equation
 (i.e. eqn. \eqref{eqn-MEalt-curlB}).

%
\subsection{Simple case: isotropic \& homogeneous}

Since trying to treat all the details of eqn. \eqref{eqn-D2owe-CX}
 correctly leads to excessively complicated expressions, 
 I will first simplify it down into the case where 
 (a) the material's reference properties ${a}_{\pPermeability}$, ${a}_{\pPermittivity}$ 
  do not vary in space, 
  and
 (b) the effective monopole current is zero (${\pKcurrentv}=0$).
Since (magnetic) monoples do not exist, 
 and all non reference properties can still be encoded in 
 ${\pPfieldv}_{\textup{\pDfield}}$, ${\pMfieldv}_{\textup{\pBfield}}$,
 or ${\pCurrentv}$, 
 this simplification does not (of itself)
 impose any approximation.
However, 
 the first simplification means that our reference properties 
 cannot incorporate any exact knowledge 
 about the spatial structure of the propagation medium\footnote{This differs
   from the situation present in the 
   complementary spatially propagated approach \cite{Kinsler-2010pra-fchhg},
   where the medium's linear temporal response typically \emph{can} be subsumed
   into the reference behaviour.
  Nevertheless, 
   in neither case can the \emph{spatial} properties be incorporated exactly 
   into the reference behaviour; 
   although approximations 
   {that assume a particular $k$ dependence can be made.}
}.
This restriction on obtaining the best possible match
 between the reference behaviour and the exact behaviour
 of the propagation medium will typically affect later approximations,
 where we assume deviations from the reference behaviour
 are small.
One additional advantage of selecting constant reference parameters 
 $\alpha_{\pPermeability}$ and $\alpha_{\pPermittivity}$ 
 is that we can {easily} replace them with matrices to allow for 
 e.g. birefringence and cross-polarization couplings.

The simplified second order wave equation 
 for changes in the displacement field ${\pDfieldv}(\Vec{r})$ 
 as it propagates forward in time is 
~
\begin{align}
  \partial_t^2
    {\pDfieldv}
&=
 -
  {\pLightspeed}^2
  {a}_{\pPermeability} 
    {a}_{\pPermittivity}
    \grad
   \cross
    \grad
   \cross
    {\pDfieldv}
\quad
 +
  {\pLightspeed}^2
  {a}_{\pPermeability} 
    \grad
   \cross
    \grad
    \cross
    {\pPfieldv}_{\textup{\pDfield}}
\nonumber
\\
&\qquad\qquad
 -
  \grad
  \cross
  \partial_t {\pMfieldv}_{\textup{\pBfield}}
 -
  \partial_t {\pCurrentv}
\label{eqn-D2owe-C3-Simple}
\\
&=
 +
  {\pLightspeed}^2
  {a}_{\pPermeability} 
    {a}_{\pPermittivity}
    \grad^2
    {\pDfieldv}
 -
  {\pLightspeed}^2
  {a}_{\pPermeability} 
    {a}_{\pPermittivity}
    \grad 
    \grad \cdot
    {\pDfieldv}
\nonumber
\\
&
\qquad\quad
 -
  {\pLightspeed}^2
  {a}_{\pPermeability} 
    \grad^2
    {\pPfieldv}_{\textup{\pDfield}}
 +
  {\pLightspeed}^2
  {a}_{\pPermeability} 
    \grad
    \grad \cdot
    {\pPfieldv}_{\textup{\pDfield}}
\nonumber
\\
&
\qquad\quad\quad
 -
  \grad
  \cross
  \partial_t {\pMfieldv}_{\textup{\pBfield}}
 -
  \partial_t {\pCurrentv}
.
\label{eqn-D2owe-C4-Simple}
\end{align}
Note that the retention of the $\grad \cdot {\pDfieldv}$
 and $\grad \cdot {\pPfieldv}_{\textup{\pDfield}}$ terms allows us to include charge effects, 
 as is needed when discussing high-power nonlinear situations 
 such as filamentation \cite{Berge-S-2009dcdsa}.

In the spatial Fourier ``wavevector'' regime, 
 we replace $\grad \rightarrow \imath \Vec{k}$, 
 so that $k^2 = \Vec{k} \cdot \Vec{k}$,
 and reorganize slightly.
The equation for changes in the displacement field ${\pDfieldv}(\Vec{k})$ 
 as it propagates forward in time
 is 
~
\begin{align}
  \partial_t^2
    {\pDfieldv}
 -
  {\pLightspeed}^2
    {a}_{\pPermeability} 
    {a}_{\pPermittivity}
  k^2
    {\pDfieldv}
&=
 -
  {\pLightspeed}^2
  {a}_{\pPermeability} 
    k^2
    {\pPfieldv}_{\textup{\pDfield}}
 -
  \partial_t {\pCurrentv}
 -
  \imath \Vec{k} 
  \cross
  \partial_t {\pMfieldv}_{\textup{\pBfield}}
\nonumber
\\
&\qquad
 -
  {\pLightspeed}^2
    {a}_{\pPermeability} 
    {a}_{\pPermittivity}
    \Vec{k} \Vec{k} \cdot {\pDfieldv}
 +
  {\pLightspeed}^2
  {a}_{\pPermeability} 
    \Vec{k} \Vec{k} \cdot {\pPfieldv}_{\textup{\pDfield}}
\label{eqn-D2owe-K-Simple}
\\
  \partial_t^2
    {\pDfieldv}
 -
  \Omega^2
    {\pDfieldv}
&=
 +
  \Omega^2
  {a}_{\pPermittivity}^{-1}
    {\pPfieldv}_{\textup{\pDfield}}
 -
  \partial_t {\pCurrentv}
 -
  \imath \Vec{k} 
  \cross
  \partial_t {\pMfieldv}_{\textup{\pBfield}}
\nonumber
\\
&\qquad
 -
  \Omega^2 k^{-2}
    \Vec{k} \Vec{k} \cdot {\pDfieldv}
 +
  \Omega^2 k^{-2}
  {a}_{\pPermittivity}^{-1}
    \Vec{k} \Vec{k} \cdot {\pPfieldv}_{\textup{\pDfield}}
,
\label{eqn-D2owe-K2-Simple}
\end{align}
 where we have defined a reference frequency $\Omega$ for the propagation, 
 based on the reference material properties, 
 i.e. 
~
\begin{align}
  \Omega^2(\Vec{k})
&=
  {\pLightspeed}^2 
  {a}_{\pPermeability} {a}_{\pPermittivity}
  k^2
.
\label{eqn-D2owe-K2-Simple-Omega}
\end{align}
We could --
 if we believed we understood the 
 spatial properties of the propagation medium sufficiently well --
 simply alter this definition 
 by incorporating an alternate $\Vec{k}$ dependence
 that mimics an assumed spatial dispersion.
However, 
 it needs to be understood that 
 such a step is distinctly \textit{ad hoc},
 since it lacks the convolutions over $\Vec{k}$ necessary for 
 any accurate representation
 of the material structure.
But, 
 that said, 
 such an approach may still have considerable practical utility; 
 and indeed is functionally equivalent to how 
 spatial dispersion {can be handled} 
 as a non-local (and hence, 
 strictly speaking, 
 a non-causal)
 process: 
 e.g. those that import a waveguide dispersion
 calculated from the (spatial) modes given by the guide's
 transverse cross-section.
To allow for this possibility, 
 in the following I will replace
 the physically rigorous magnitude $k$ argument for $\Omega$
 with $\Vec{k}$, 
 its vector counterpart\footnote{In some cases, 
   depending on the approximate model chosen, 
   this new $\Omega(\Vec{k})$ definition may also need to be convolved
   with other parts of the expression, 
   e.g. $\Omega^2(k) {\pPfieldv}_{\textup{\pDfield}}(\Vec{k})
     \rightarrow \Omega^2(\Vec{k}) \convol {\pPfieldv}_{\textup{\pDfield}}(\Vec{k})$, 
   although such an additional complication is likely to 
   prevent us obtaining much conceptual or mathematical benefit.
  Consequently, 
   in what follows I do not allow for such convolutions.}.
This allows for not only more complicated assumed spatial dispersions, 
 but also additional orientation dependence.
However,
 these benefits (only) arise as a result of approximation.
Strictly speaking, 
 both on the grounds covered above, 
 and ones taking stricter view of causality \cite{Kinsler-2014arXiv-negfreq}, 
 {many simplistic notions of spatial dispersion
 involving adding some convenient  $k$ dependence
 is incompatible with locality\footnote{{Of course, 
  it is possible to treat spatial dispersion in a properly causal way; 
  a notable example being
  the hydrodynamic plasmon model \cite{Ciraci-PS-2013cphc}.
  This takes propagation in the electron fluid into account, 
  giving rise to an additional $k^2$ dependance.}}.}

Unlike for the spatially propagated scheme 
 in \cite{Kinsler-2010pra-fchhg}, 
 we have not included temporal response of the propagation medium 
 in our definition of 
 the directional fields,
 since it is incompatible with our temporal propagation.
To include this
 we must model it explicitly as is done in 
 Maxwell solvers such as FDTD \cite{Yee-1966tap,Oskooi-RIBJJ-2010cpc}; 
 or approximate it as if it were a spatial effect
 (see e.g. \cite{Carter-1995pra}, 
 or the discussions
 in the later section \ref{S-dispersion}. 

%
\section{Factorization}
\label{S-factorisation}

{I now factorize the second order wave equation for $\pDfieldv$
 \cite{Ferrando-ZCBM-2005pre,Kinsler-2010pra-fchhg}.
This process neatly avoids 
 some of the approximations necessary in traditional approaches, 
 and takes its name from the fact that 
 the LHS of eqn. \eqref{eqn-D2owe-C4-Simple} 
 or \eqref{eqn-D2owe-K2-Simple} 
 is a simple difference of squares which might be factorized.
Indeed,
 this was done by Blow and Wood in 1989 \cite{Blow-W-1989jqe}, 
 albeit in a rather ad hoc --
 but nevertheless effective --
 fashion.
Since the factors are just of the form $(\partial_t \mp \imath k)$, 
 taken individually they each look like 
 a simple forward (or backward) directed first order wave equation.
The factorization method therefore allows us to define
 a pair of counter-propagating Greens functions, 
 which divides the original second order wave equation
 into a pair of oppositely directed first order wave equations
 that are coupled together.
In addition, 
 it allows us to straightforwardly compare the exact bi-directional
 and approximate uni-directional theories term by term, 
 which is in distinct contrast to other approaches, 
 where the backward or unwanted parts
 are approximated away piecemeal,
 and may not be suitable for direct comparison.
A short summary of the factorization method, 
 adapted to this new context from earlier work \cite{Kinsler-2010pra-fchhg}, 
 is given in appendix \ref{S-factorize}.}

{An important point to remember is that 
 the choice of ${a}_{\pPermittivity}, {a}_{\pPermeability}$ 
 and therefore ${\Omega}(\Vec{k})$
 in eqn. \eqref{eqn-D2owe-K2-Simple} 
 defines the specific Greens functions used.
This means that it 
 also defines the basis 
 upon which we will then propagate the displacement field ${\pDfieldv}$.}

{In this section and the next, 
 the derivation will closely follow the proceedure, 
 language, 
 and terminology used 
 in earlier work \cite{Kinsler-2010pra-fchhg}.
However, 
 despite the mathematical similarities, 
 it is important not to lose sight of the distinct physical differences.
That previous work focussed on propagation of optical fields
 along a spatial axis, 
 whereas here I instead consider propagation in time.
Since my aim is to compare and contrast the two approaches, 
 as discussed at length in sections \ref{S-dispersion} and \ref{S-nls},
 it is useful to enable comparisons  not only 
 at the endpoint of the derivation, 
 but also at each step along the way.}

%
\subsection{Bi-directional wave equations}

A pair of bi-directional wave equations suggests 
 similarly bi-directional fields, 
 so I split the electric displacement field 
 into forward (${\pDfieldv}^+$) and backward (${\pDfieldv}^-$) directed parts, 
 with ${\pDfieldv} = {\pDfieldv}^+ + {\pDfieldv}^-$.
{In the following,} 
 remember that ${\pPfieldv}_{\textup{\pDfield}}$, 
 ${\pMfieldv}_{\textup{\pBfield}}$, etc
 all depend on the full field ${\pDfieldv}$, 
 and \emph{not} only one partial field (e.g. only ${\pDfieldv}^+$).
This is an important point since we see that they then
 drive both {the} forward and backward evolution equations equally.

The coupled bi-directional first order wave equations
 for the directed fields ${\pDfieldv}^\pm$ are
  \emph{propagated} forward in time 
 whilst being \emph{evolved} forward and backward in space are
~
\begin{align}
    \partial_t
  {\pDfieldv}^\pm(\Vec{k})
&=
 \pm
    \imath
    \Omega(\Vec{k})
  {\pDfieldv}^\pm(\Vec{k})
 ~~
 \pm
  \frac{\imath \Omega^2(\Vec{k}) {a}_{\pPermittivity}^{-1}}
       {2 \Omega(\Vec{k})}
  {\pPfieldv}_{\textup{\pDfield}}({\pDfieldv},t;\Vec{k})
\nonumber
\\
& \quad~~
 \mp
  \frac{\dot{{\pCurrentv}}(t;\Vec{k})}
       {2 \Omega(\Vec{k})}
 ~
 \mp
  \frac{\imath }
       {2 \Omega(\Vec{k})}
  \Vec{k}  \cross
  \dot{{\pMfieldv}}_{\textup{\pBfield}}({\pDfieldv},t;\Vec{k})
\nonumber
\\
&\qquad
 \mp
  \frac{\imath \Omega^2(\Vec{k}) }
       {2 \Omega(\Vec{k}) k^2}
  \Vec{k} \Vec{k} 
  \cdot
  {\pDfieldv}(\Vec{k})
 ~
 \pm
  \frac{\imath \Omega^2(\Vec{k}) {a}_{\pPermittivity}^{-1}}
       {2 \Omega(\Vec{k}) k^2}
  \Vec{k} \Vec{k} 
  \cdot
  {\pPfieldv}_{\textup{\pDfield}}({\pDfieldv},t;\Vec{k})
\\
    \partial_t
  {\pDfieldv}^\pm(\Vec{k})
&=
 \pm
    \imath
    \Omega(\Vec{k})
  {\pDfieldv}^\pm(\Vec{k})
 ~~
 \pm
  \frac{\imath \Omega(\Vec{k}) }
       {2 {a}_{\pPermittivity}}
  {\pPfieldv}_{\textup{\pDfield}}({\pDfieldv},t;\Vec{k})
\nonumber
\\
& \quad~~
 \mp
  \frac{\dot{{\pCurrentv}}(t;\Vec{k})}
       {2 \Omega(\Vec{k})}
 ~
 \mp
  \frac{\imath }
       {2 \Omega(\Vec{k})}
  \Vec{k}  \cross
  \dot{{\pMfieldv}}_{\textup{\pBfield}}({\pDfieldv},t;\Vec{k})
\nonumber
\\
&\qquad
 \mp
  \frac{\imath \Omega(\Vec{k}) }
       {2}
  \frac{\Vec{k} \Vec{k}}
       {k^2}
  \cdot 
     {\pDfieldv}(\Vec{k})
 ~
 \pm
  \frac{\imath \Omega(\Vec{k})}
       {2 {a}_{\pPermittivity}}
  \frac{\Vec{k} \Vec{k}}
       {k^2}
  \cdot
    {\pPfieldv}_{\textup{\pDfield}}({\pDfieldv},t;\Vec{k})
\label{eqn-bi-dtD}
\end{align}
Here I have replaced the partial time derivatives 
 on the RHS with over-dots, 
 this being to emphasise that if our expression is to remain 
 strictly causal \cite{Kinsler-2011ejp}, 
 then our models for current ${\pCurrentv}$ 
 and magnetization ${\pMfieldv}_{\textup{\pBfield}}$
 {should} return their \emph{time-derivatives} as explicit functions
 of known quantities (here, typically ${\pDfieldv}$ or $t$).
{For example, 
 a current model specification such as ${\pCurrentv}=\sigma{\pDfieldv}$
 would mean that $\dot{{\pCurrentv}}=\partial_t {\pDfieldv}$, 
 giving a $\partial_t {\pDfieldv}$ on both sides.}

Note that factorized equations can be rebuilt
 into a single second-order equation by taking the sum and difference, 
 then substituting {one into the other} with the assistance
 of a time derivative
 (see section IV.B of \cite{Ferrando-ZCBM-2005pre}).

{A remaining clarification required is to understand the meaning
 of ``forward'' and ``backward'' in the context
 of three dimensional space --
 this being less straightforward than in the spatially propagated case, 
 where one spatial axis is selected
 for propagation\footnote{Typically a cartesian axis, 
 although other choices are possible \cite{Kinsler-2012arxiv-fbrad}.}, 
 and forward and backward refers to time $t$.
Here, 
 we should understand ``forward'' and ``backward'' as being
 with respect to a given choice of $\Vec{k}$.}

%
\subsection{Propagation, evolution, and directed fields}
\label{S-factorisation-propevodir}

{Since we have chosen to 
 enforce \emph{propagation} towards later times $t$
 on our solutions of the wave equations},
 the fields ${\pDfieldv}^\pm(t)$
 are \emph{directed} forwards and backwards in space; 
 these fields then \emph{evolve} 
 forwards and/or backwards in space as $t$ increases.
I use this terminology 
 (propagated, directed, evolved)
 throughout this paper
 to mean these three specific things.
{This usage is consistent with that
 used in the alternative spatially propagated approach
 \cite{Kinsler-2010pra-fchhg}, 
 but in that case the fields $\pEfieldv^\pm(z)$
 are directed in time, 
 and evolve forwards and/or backwards in time as $z$ increases.}


{The wave equation eqn. \eqref{eqn-bi-dtD}
 which evolves the directed fields ${\pDfieldv}^\pm$
 as they propagate forward in $t$, 
 has two types of terms on the RHS:
 and I call these the ``reference'' 
 and ``residual'' parts \cite{Kinsler-2009pra,Kinsler-2010pra-fchhg}.}

\emph{Reference evolution}
 (or \emph{``underlying evolution''})
 is that given by $\pm \imath \Omega (\Vec{k} ){\pDfieldv}^\pm$ term, 
 and is determined by our chosen 
 ${a}_{\pPermittivity}$ and ${a}_{\pPermeability}$, 
 and any additional ad hoc refinements.
By itself, 
 it would describe an ordinary oscilliatory evolution where the 
 field oscillations in ${\pDfieldv}^\pm(\Vec{r})$
 would move forward ($+$) or backwards ($-$)
 in space.
This is analogous to the choice of reference when 
 constructing directional fields \cite{Kinsler-RN-2005pra}, 
 but here is done for temporal 
 rather than spatial propagation \cite{Kinsler-2012arXiv-fbacou}.

\emph{Residual evolution}
 accounts for the discrepancy between the true evolution
 and the underlying/reference evolution.
{It consists of all parts
 of the material response \emph{not} 
 included in ${a}_{\pPermittivity}, {a}_{\pPermeability}$ (and hence $\Omega(\Vec{k})$); 
 i.e. it contains all of the 
 rest of the terms on the RHS of eqn. \eqref{eqn-bi-dtD}.
These will usually consist of any non-linear polarization,  
 or spatially or orientationally dependent linear terms; 
 they are analogous to the correction terms 
 used in methods based on directional fields.
Alternatively, 
 such contributions are called ``source'' terms
 by Ferrando et al. \cite{Ferrando-ZCBM-2005pre}.
Generally we will hope they only provide a weak perturbation, 
 so that we might make the (desirable) uni-directional approximation
 discussed later; 
 but the factorization procedure itself 
 is valid \emph{regardless} of their strength.}

%
\subsection{Reference evolution: 
              choice of $\Omega$ and the resulting ${\pDfield}^\pm$}
\label{S-factorisation-beta}

{I now consider how our choice of $\Omega$ will affect the 
 relative sizes of the forward and backward directed 
 ${\pDfieldv}^+$ and ${\pDfieldv}^-$.
I therefore now consider
 the example of a simple medium in which the field
 is known to propagate with frequency $\omega$; 
 but instead we choose a reference evolution 
 determined by a frequency $\Omega$
 that is different from $\omega$.}
E.g., 
 for a linear isotropic medium we could exactly define
 $\omega^2 = \Omega^2 + \Delta^2$; 
 but in general we would just have some residual (source) term $\mathscr{Q}$.
{As a consequence,
 the definitions of forward and backward \emph{directed} fields
 do not correspond exactly to what the wave equation
 actually will \emph{evolve} forward and backward,
 as we \emph{propagate} towards later times.}

The second order wave equation is
 $(\partial_t^2+\Omega^2) {\pDfieldv} = - \mathscr{Q}$, 
 which in the linear case has $\mathscr{Q}=\Delta^2 {\pDfieldv}$, 
 so that $(\partial_t^2+\omega^2) {\pDfieldv} = 0$.
The factorization
 in terms of $\Omega$ is then
~
\begin{align}
  \partial_t
  {\pDfieldv}^\pm
&=
 \pm
  \imath \Omega {\pDfieldv}^\pm
 \pm
  \frac{\imath \mathscr{Q}}{2 \Omega}
.
\label{eqn-choosebeta-factor}
\end{align}
Now if we choose the situation
 where our field ${\pDfieldv}$ only evolves forward, 
 we know that ${\pDfieldv} = {\pDfieldv}_0 \exp [\imath \omega t]$.
Consequently ${\pDfieldv}^\pm$ must have matching oscillations: 
 i.e. ${\pDfieldv}^\pm = {\pDfieldv}_0^\pm \exp [\imath \omega t]$,
 even though ${\pDfieldv}^-$ is directed backwards.
Substituting these 
 into eqn. \eqref{eqn-choosebeta-factor} gives
~
\begin{align}
  {\pDfieldv}_0^-
&=
  \frac{\Omega - \omega}
       {\Omega + \omega}
  {\pDfieldv}_0^+
.
\label{eqn-choosebeta-EmEp}
\end{align}
{This tells us how much ${\pDfieldv}^-$
 we need to combine with ${\pDfieldv}^+$
 so that our pulse evolves forward; 
 the ${\pDfieldv}^-$ is strongly coupled to ${\pDfieldv}^+$, 
 and so will be dragged forward against its usual preference.}
This interdependence of ${\pDfieldv}^\pm$ is generic -- 
 no matter what the origin of the discrepancy between $\Omega$
 and the true evolution of the field
 (i.e. the residual/source terms such as 
  mismatched linear behaviour, nonlinearity, etc):  
 some non-zero backward directed field ${\pDfieldv}^-$ must exist
 but still evolve forwards with ${\pDfieldv}^+$.
{Comparable behaviour is also} seen in the directional fields
 approach of Kinsler et al. \cite{Kinsler-RN-2005pra}.

{Typically we will hope that this residual ${\pDfieldv}^-$ contribution
 is small enough so as to be negligible.}
If we assume ${\pDfieldv}^- \simeq 0$, 
 then we find that $\omega \simeq \Omega + \Delta^2/2\Omega$, 
 which is just the expansion of $\omega = (\Omega^2 + \Delta^2)^{1/2}$
 to first order in $\Delta^2/\Omega^2$.
Following this, 
 we find that eqn. \eqref{eqn-choosebeta-EmEp} then says that 
 ${\pDfieldv}_0^- \simeq (\Delta^2/4\Omega^2) {\pDfieldv}_0^+$, 
 which {now provides} us with the scale
 on which ${\pDfieldv}^-$ can be {ignored.
Apart from the simple} (linear) case where we know $\Delta^2$, 
 the true frequency $\omega$ might be difficult to determine, 
 and in nonlinear propagation
 any local estimate of the frequency can 
 {change during propagation}.

{Further,} 
 if we choose $\Omega=\Omega(k)$ with a wave-vector dependence, 
 then we see that the source-like terms
 inherit that dispersion; 
 and this is almost inevitable, 
 since typically $\Omega \sim c k$.
This means, 
 for example, 
 that even if we start with model of nonlinear polarization  
 without any $k$ dependence, 
 our factorized equations will have a $k$ dependence
 on the nonlinear terms.
This is the complementary process 
 to what happens in the spatially propagated 
 {approaches of Kinsler et al. 
  \cite{Kinsler-RN-2005pra,Kinsler-2010pra-fchhg}, 
 where choosing a dispersive reference has consquences for the 
 behaviour of the correction terms.}

%
\section{Uni-directional wave equations}
\label{S-firstorder}

{We can now make a single,
 well defined class of approximation, 
 and simplify the
 exact coupled bi-directional evolution of ${\pDfieldv}^\pm$ down to  
 just one uni-directional 
 first order wave equation.
As with the complementary spatially propagated theory \cite{Kinsler-2010pra-fchhg},
 this does not require a moving frame, 
 a smooth envelope, 
 or to assume inconvenient second order derivatives 
 are somehow negligible,
 as was necessary in traditional treatments.
The approximation assumes that the residual terms are weak
 in comparison to the (reference) $\pm \imath \Omega {\pDfieldv}$ term --
 e.g. weak nonlinearity or orientation dependence.
This means that I can assert that if we start with ${\pDfieldv}^-=0$, 
 then ${\pDfieldv}^-$ will stay negligible.
In this context, 
 ``weak'' means that no significant change in the backward field 
 is generated in a time shorter than one period
 (``slow evolution'');}
 and that small effects do not build up gradually over propagation times 
 of many periods (``no accumulation'').

\emph{Slow evolution} 
 is where the size of the residual terms 
 is much smaller than that of the underlying linear evolution --
 i.e. smaller than $\Omega {\pDfieldv}$.
This allows us to write down straightforward inequalities
 which need to be satisfied.
It is important to note the close relationship between these and 
 a good choice of $\Omega$, 
 as discussed in subsection \ref{S-factorisation-beta}.
If $\Omega$ is not a good enough match, 
 there always be significant contributions from both
 forward and backward directed fields; 
 and even if nothing ends up \emph{evolving} backwards, 
 an ignored backward directed field will result
 in (e.g.) miscalculated nonlinear effects, 
 since the total field ${\pDfieldv} = {\pDfieldv}^+ + {\pDfieldv}^-$ will be wrong.

\emph{No accumulation} 
 occurs when the evolution of any backward directed field ${\pDfieldv}^-$
 is dominated by its coupling via the residual terms
 to the forward directed field ${\pDfieldv}^+$; 
 and not by its preferred underlying backward evolution.
No accumulation means that forward evolving field components
 do not couple to field components that evolve backwards; 
 this the typical behaviour since the phase mismatch 
 between forward evolving and backward evolving components
 is $\sim 2\Omega$; 
 in essence it is comparable
 to the common rotating wave approximation (RWA).
This rapid relative oscillation
 means that backward evolving components never accumulate, 
 as each new addition will be out of phase with the previous one; 
 it is not quite a ``no reflection'' approximation, 
 but one that asserts that the many micro-reflections
 do not combine to produce something significant.
An estimate of the terms required to break this approximation
 are given in appendix \ref{S-RWA}; 
 generally speaking this is a much more robust approximation
 than the slow evolution one.
Of course, 
 a periodic temporal modulation of the medium 
 will give periodic residual terms, 
 and these can be engineered to force phase matching.
In spatially propagated analyses, 
 where we talk of spatial (wavevector) phase matching
 and not this temporal (frequency) phase matching
 \cite{Kinsler-NT-2006-phnlo}, 
 this would be a periodicity based on a relatively 
 small phase mismatch 
 (see e.g. quasi phase matching in Boyd \cite{Boyd-NLO}); 
 but might even go as far as matching the backward wave 
 (see e.g. \cite{Harris-1966apl}).

{Note however that small forward perturbations
 from the residual terms \emph{can} accumulate on the 
 forward evolving field components, 
 as indeed can backward perturbations accumulate on the 
 backward evolving field components.
Despite the fact that the residual terms acting on
 the forward and backward field evolutions
 are the same size,}
 forward evolving components of the residuals 
 can accumulate on the forward evolving field
 because they are phase matched;  
 whereas backward residuals are not, 
 and rapidly average to zero.

%
\subsection{Residual terms}

The handling and criteria for different types
 of residual terms has been discussed already \cite{Kinsler-2010pra-fchhg}
 for the spatially propagated case.
While it would be straightforward to repeat that level of detail
 here, 
 the basic methodology is virtually identical except for
 the different roles played by the $t$ and $\Vec{r}$ arguments.
Accordingly I will 
 consider only the most commonly considered residuals --
 the dielectric scalar and vector terms
 as well as currents.
Those interested in magnetic or magneto-electric effects
 can adapt the process to their own systems in a similar manner.

The relevant total polarization ${\pPfieldv}_{\textup{\pDfield}}$ 
 can then be decomposed into pieces
 before seeing how each might satisfy the slow evolution criteria.
Thus we write 
~
\begin{align}
  {\pPfieldv}_{\textup{\pDfield}}({\pDfieldv},\Vec{r},t)
&=
  {a}_{\pPermittivity} ({\pDfieldv},\Vec{r},t)
  {\pDfieldv}(\Vec{r},t)
 +
  \Vec{U}_{\textup{\pDfield}} ({\pDfieldv},\Vec{r},t)
\nonumber
\\
&=
  {a}_L (\Vec{r},t)
   \ast 
  {\pDfieldv}(\Vec{r},t)
 +
  {a}_N ({\pDfieldv},\Vec{r},t)
   \ast 
  {\pDfieldv}(\Vec{r},t)
\nonumber
\\
&\qquad
 +
  \Vec{U}_L (\Vec{r},t) \ast {\pDfieldv}(\Vec{r},t)
 +
  \Vec{U}_N ({\pDfieldv},\Vec{r},t)
.
\end{align}
{The scalar part is} represented by ${a}_{\pPermittivity}$, 
 and might contain (non-reference) linear parts and time response (${a}_L$), 
 perhaps a convolution (``$\ast$'') over the past, 
 or angle dependence; 
 it might also 
 {include} nonlinear contributions ${a}_N$ such as a 
 third order nonlinearity
 with ${a}_N {\pDfieldv} \propto ({\pDfieldv} \cdot {\pDfieldv}) {\pDfieldv}$.
The vector part $\Vec{U}_{\textup{\pDfield}}$, 
 if present, 
 could be due to {something like} a second order nonlinearity, 
 which couples the ordinary and extra-ordinary field polarizations.
{This description of the material parameters
 is not restrictive,
 might e.g. include any order of nonlinearity.}

%
\subsection{Slow evolution?}

{Having categorized the relevant residual terms, 
 it now remains to treat each one individually
 and determine the conditions under which 
 the oppositely directed field can nevertheless remain} negligible:
 i.e., for ${\pDfieldv}^\pm$, we have that ${\pDfieldv}^\mp\simeq 0$.
 where the scalar ${a}_L$ contains the
 linear response of the material that is both isotropic 
 and lossless (or gain-less); 
 since here it is a time-response function, 
 it is convolved with the electric displacement field ${\pDfieldv}$.
Note that the field vector ${\pDfieldv}$,
 and indeed the material parameter ${a}_L$
 are all functions of time $t$ and space $\Vec{r} = (x,y,z)$; 
 the polarization ${\pPfieldv}_{\textup{\pDfield}}$ and its components
 ${a}_{\pPermittivity}$, $\Vec{U}_{\pPermittivity}$
 are a functions of time $t$, space $\Vec{r}$, 
 and the field ${\pDfieldv}$.

Below,
 {the field vector ${\pDfieldv}$ will be split into components}
 ${\pDfield}_i$ and ${\pDfield}_i^\pm$, 
 with ${\pDfieldv}^+ + {\pDfieldv}^- 
  = {\pDfieldv} \equiv ({\pDfield}_x,{\pDfield}_y,{\pDfield}_z)$
 and $i \in \{x,y,z\}$. 
{Like we also have
 wavevector components $k_i$ from $\Vec{k} = (k_x, k_y, k_z)$, 
 where $k_\perp^2 = k_x^2 + k_y^2$;
 and further,}
 $k_j$ is also used as a substitute symbol
 to represent any one of $k_x$, $k_y$, or $k_z$.

Any corrections due to imperfect choice of $\Omega$
 will appear in the ${\pPfieldv}_{\textup{\pDfield}}$ polarization correction term.
However, 
 unlike in the spatially propagated case, 
 there is no diffraction-like term --
 here, 
 that is quite naturally part of the reference behaviour.

~\\
\emph{Firstly,} 
 {we consider the}
 scalar polarization terms ${a}$,
 which {could} be either linear (${a}_L(k)$) 
 or nonlinear (${a}_N({\pDfieldv},k)$).
These might {represent}
 e.g. the time-response of the medium (and hence its dispersion), 
 or be some nonlinear response such as the 
 third-order Kerr nonlinearity.
Since these are locally correlated in space, 
 in the wavevector picture they include convolutions over $k$.
{The criterion therefore is}
~
\begin{align}
  \frac{\imath \Omega ~{a} \convol \left| {\pDfield}_i^+ + {\pDfield}_i^- \right|/ 2 }
       {\imath \Omega \left| {\pDfield}_i^\pm \right| }
&\simeq
  \frac{{a}\convol \left| {\pDfield}_i^\pm \right| / 2}
       {\left| {\pDfield}_i^\pm \right| }
&\simeq
  \frac{{a}}
       {2}
\quad
\ll
  1
.
\label{eqn-uni-scalarapprox}
\end{align}
{In the important linear case, 
 ${a} \equiv {a}_L$ is independent of ${\pDfieldv}$, 
 which means that the pulse properties play no role, 
 and only the material parameters are constrained.}
In the nonlinear case, 
 {there is a further constraint, 
 which is on the peak intensity of the pulse,
 but still there are no}
 smoothness assumptions or bandwidth restrictions.

~\\
\emph{Secondly,}
 {we consider the}
 linear and nonlinear vector terms from $\Vec{U}$.
{This criterion is similar to the 
 the scalar case} in eqn. \eqref{eqn-uni-scalarapprox},
 but with $\Vec{U}$ replacing $\phi {\pDfieldv}$.
Thus for $i \in \{x,y,z\}$, 
 we can write down constraints for each component $U_i$,
 which are
~
\begin{align}
   \imath \Omega \left| {U}_{i} \right| / 2
& \ll
  \imath \Omega \left| {\pDfield}_i^\pm \right|
& \Longrightarrow \qquad
  \left| {U}_{i} \right|
& \ll
  2 \left| {\pDfield}_i^\pm \right|
.
\label{eqn-uni-vectorapprox}
\end{align}
In the linear case, 
 $\Vec{U} \equiv \Vec{U}_L$, 
 {and the linear relationship between $\Vec{U}_L$ and ${\pDfieldv}$
 again means that
 the} criterion only constrains the material parameters
 contained in $\Vec{U}_L$,
 not the field profile or amplitude.
The nonlinear case is similar to the scalar one, 
 and the peak pulse intensity is restricted; 
 e.g. for a $\chi^{(2)}$ medium, 
 $|\Vec{U}_N| \sim \chi^{(2)} |{\pDfieldv}|$.

{An additional complication 
 that occurs in this vector case}
 is that a field consisting
 of only one field polarization component (e.g. ${\pDfield}_x^+$)
 may induce a driving in the orthogonal 
 (and initially zero) components (e.g. ${\pDfield}_y^\pm$).
{This means that}
 both ${\pDfield}_y^\pm$ fields will be driven with the same strength, 
 so that it is far from obvious that we can set ${\pDfield}_y^-$ to zero, 
 but still keep the ${\pDfield}_y^+$ without being inconsistent.
{However, 
 it has already been noted that 
 phase matching ensures that} forward residuals accumulate, 
 whilst the non-matched backward residuals are subject to the RWA, 
 and become negligible:
 hence we can still rely on eqn. \eqref{eqn-uni-vectorapprox}, 
 albeit under caution.

~\\
\emph{Thirdly,}
 we have the residual terms indicated by the presence 
 of time-varying currents.
These are
~
\begin{align}
  \left| \dot{J}_i  \right|
& \ll
   2 \Omega^2 \left| {\pDfield}_i^\pm \right|
.
\label{eqn-uni-current}
\end{align}

~\\
\emph{Fourthly,}
 we have the divergence terms $\Vec{k} \Vec{k} \cdot {\pDfieldv}$, 
 $\Vec{k} \Vec{k} \cdot {\pPfieldv}_{\textup{\pDfield}}$, 
 which amount to charge density contributions
 either from ${\pDfieldv}$ or ${\pPfieldv}_{\textup{\pDfield}}$, 
 with $\sigma_P(k) = {a}_{\pPermittivity}^{-1} \Vec{k} \cdot {\pPfieldv}_{\textup{\pDfield}}$ 
 and $\sigma_{\textup{\pDfield}}(k) =  \Vec{k} \cdot {\pDfieldv}$.
Thus if the wavevector is oriented along a unit vector $\Vec{u}$, 
 then
~
\begin{align}
  \imath u_i \Omega \left| \sigma \right| / 2k
& \ll
   \imath \Omega \left| {\pDfield}_i^\pm \right|
\nonumber
\\
  \sigma u_i & \ll 2 k \left| {\pDfield}_i^\pm \right|
.
\label{eqn-uni-divergence}
\end{align}
 is the relevant criterion.

~\\
\emph{To summarize}, 
 these various criteria assert
 that modulations away from the reference evolution must be weak.
Notably, 
 weak nonlinearity is invariably guaranteed by material damage thresholds
 \cite{Kinsler-2007josab}.
However, 
 and in contrast to the usually favoured 
 (but strictly speaking less causal)
 spatially propagated methods, 
 the lack of any rigorous way to incorporate dispersion into
 the reference frequency $\Omega(k)$
 makes it harder to satisfy the criteria 
 for the ordinary linear response.

%
\subsection{Uni-directional equation for ${\pDfieldv}^+$}

{In a situation in which 
 all of the above slow-evolution criteria hold, 
 we can be sure that
 the backward directed field ${\pDfieldv}^-$
 is only driven by a negligible amount, 
 and if the no-accumulation condition also holds, 
 then neither will there be any build up of 
 backward evolving contributions to the field.}
Consequently, 
 we can be sure that an initially negligible ${\pDfieldv}^-$ remains so, 
 and eqn.\eqref{eqn-bi-dtD} becomes
~
\begin{align}
    \partial_t
  {\pDfieldv}^+(\Vec{k})
&=
 +
    \imath
    \Omega(\Vec{k})
  {\pDfieldv}^+(\Vec{k})
 ~~
 +
  \frac{\imath \Omega(\Vec{k})}
       {2 {a}_{\pPermittivity}}
  {\pPfieldv}_{\textup{\pDfield}}
  (
    {\pDfieldv}^+, t; \Vec{k}
  )
\nonumber
\\
&~\quad
 -
  \frac{\dot{{\pCurrentv}}(t;\Vec{k})}
       {2 \Omega(\Vec{k})}
 ~~
 -
  \frac{\imath}
       {2 \Omega(\Vec{k})}
  \Vec{k} \cross \dot{{\pMfieldv}}_{\textup{\pBfield}} 
  (
    {\pDfieldv}^+, t; \Vec{k}
  )
\nonumber
\\
& ~~\quad
 -
  \frac{\imath \Omega(\Vec{k})}
       {2}
  \frac{\Vec{k}}
       {k^2}
   \Vec{k}
   \cdot
   {\pDfieldv}^+(\Vec{k})
  ~~
 +
  \frac{\imath \Omega(\Vec{k})}
       {2 {a}_{\pPermittivity}}
  \frac{\Vec{k}}
       {k^2}
   \Vec{k}
   \cdot
  {\pPfieldv}_{\textup{\pDfield}}
  (
    {\pDfieldv}^+, t; \Vec{k}
  )   
.
\label{eqn-uni-dtD}
\end{align}
where now the polarization ${\pPfieldv}_{\textup{\pDfield}}$, 
 magnetization,
 and divergence are solely dependent 
 on the forward directed field ${\pDfieldv}^+$.

%
\subsection{Modifications}
\label{S-modifications}

Just as for the comparable spatially propagated derivation
 \cite{Kinsler-2010pra-fchhg}
 we might also apply some of the strategies used in other approaches
 to get a more tractable and/or simpler evolution equation.
Unlike in ``traditional'' derivations
 \cite{Brabec-K-1997prl,Kinsler-N-2003pra,
       Geissler-TSSKB-1999prl,Scalora-C-1994oc}, 
 none of these are required, 
 but they nevertheless may be useful.

Briefly, 
 a \emph{co-moving frame} might be added, 
 using $z'=z+t v_f$.
This is a simple linear process that causes no extra complications;
 the leading RHS $\imath c {\pDfield}^+$ term is replaced
 by $\imath (\pLightspeed \mp v) {\pDfield}^\pm$.
Note that setting $\pLightspeed=v$ will freeze the phase velocity
 of the pulse, 
 not the group velocity.
Also, 
 a \emph{carrier-envelope} separation 
 could be implemented using
 ${\pDfield}(z) = A(z) \exp [\imath (\omega_1 t - k_1 z)] 
        + A^*(z) \exp [-\imath (\omega_1 t - k_1 z)] $
 defining the envelope $A(z)$ 
 with respect to wavevector $k_1$ and carrier frequency $\omega_1$.
Multiple envelopes centred at different wavevectors
 and carrier frequencies ($k_i$, $\omega_i$) could also be used
 to split the field up.
This is typically done when the field principally comprises
 a number of distinct narrowband components
 (e.g. 
  \cite{Kinsler-N-2004pra,Kinsler-N-2005pra,Boyd-NLO}).
 The wave equation can then be separated into one equation for 
 each piece, 
 coupled by the appropriate wavevector-matched polarization terms
 (c.f. \cite{Kinsler-NT-2006-phnlo}).

%
\section{Dispersion}
\label{S-dispersion}

{The distinct contrast between the physical meaning of this
 temporally propagated wave equation
 and spatially propagated ones
 \cite{Boyd-NLO,Blow-W-1989jqe,Kinsler-2010pra-fchhg}
 now gives us an opportunity to consider the respective roles
 of temporal and spatial dispersion.
This is because the temporally propagated picture
 holds all spatial information at any given moment in time, 
 and so is a natural arena in which to treat spatial dispersion; 
 whereas the spatially propagated picture, 
 as we know, 
 holds all temporal information at any given point in space,
 and so is a natural arena in which to treat temporal dispersion.
Nevertheless, 
 the comparison is not as straightforward as we would like, 
 so in what follows I restrict the discussion 
 to the context of the simple slab waveguide.
}


{A dielectric slab waveguide
 is a high index planar sheet (the core)
 of high index material
 clad on both sides by semi-infinite volumes of lower index material.
Here the waveguide is taken to have}
 thickness $d$ in the perpendicular $x$ direction, 
 propagation in the $z$ direction, 
 and with core and cladding permittivities ${\pPermittivity}_1$ and ${\pPermittivityVac}$.
In the continuous wave (CW) limit, 
 the spatial structure of the waveguide 
 gives rise to frequency dependent properties
 that are usually called ``spatial dispersion''.
Even for this simple slab design,
 {the matching of the boundary conditions
 between core and cladding regions gives}
 the dispersion relation 
 a non trivial form; 
 in this case it is given by the solution to 
 a transcendental equation \cite{TG-FiberOptEss}.
Typically, 
 this is written in a way implying we want to calculate $k_x$ and $\beta=k_z$
 from $\omega$,
 i.e.
 for the transverse electric (TE) field modes we have
~
\begin{align}
  T(k_x d) 
&=
  k_x^{-1} 
  \sqrt{\omega^2 {\pPermeabilityVac} 
    \left(
      {\pPermittivity}_1 - {\pPermittivity}_2 
    \right)
   -
    k_x^2
  }
,\label{eq-example-zeq}
\\
\textrm{where} \quad
  \beta^2(\omega) 
&=
  \omega^2 {\pPermeabilityVac} {\pPermittivity}_1 - k_x^2
,
\label{eq-example-zk}
\end{align}
 and $T(k_x d)$ is either $\tan(k_x d)$ or $-\cot(k_x d)$.

{However, 
 we can now notice that this expression 
 looks like one where we know know $\omega$, 
 and from that have to calculate the matching wavevector properties --
 i.e.
 that this is an implicitly spatially propagated representation.
Therefore, 
 for the temporally propagated case that is the focus of this paper, 
 where $\Vec{k}$ is the known quantity, 
 eqn. \eqref{eq-example-zk} should be rewritten.
Accordingly we use
 $k_z$ for $\beta$
 (since wavevector is no longer a propagation reference), 
 and $\Omega$ for $\omega$ 
   (since frequency now \emph{is} a propagation reference);
 thus we have}
~
\begin{align}
  {\pLightspeed}^2 \Omega^2(k_x)
&=
  k_x^2
  \frac{               {\pPermittivity}_1}
       { \left( {\pPermittivity}_1 - {\pPermittivity}_2 \right)   }
  \left[ 
    T^2(k_x d) + 1
  \right]
.
\label{eq-example-teq}
\\
  \textrm{where}\quad
  k_z^2 
&=
  {\pLightspeed}^2 \Omega^2(k_x) - k_x^2
.
\end{align}
Here we see that the natural (reference) propagation frequency 
 is defined solely by the transverse wavevector $k_x$, 
 as is the matching evolution wavevector $k_z$.
While apparently simpler than the traditional form, 
 since $\Omega$ is directly given by $k_x$, 
 we usually want to neglect any direct knowledge
 of the transverse properties.
This means that what we really want is not $\Omega(k_x)$,
 but $\Omega(k_z)$,
 the inverse of the traditionally preferred $\beta(\omega)$.

Despite these complications, 
 which are in any case rather similar to those we see in the traditional
 spatially propagated picture, 
 we can still find the same waveguide modes and their dispersions, 
 but this time given as functions of $k_z$, 
 not $\omega$.
{To simplify the presentation, 
 I now assume that
 for some $n$-th mode of the waveguide, 
 we can expand about some central evolution wavelength $k_{zn}$, 
 which corrresponds to a reference frequency $\Omega_n = \Omega(k_{zn})$.
This means we can write}
~
\begin{align}
  \Omega(k_z)
&=
  \Omega_n
 + 
  \sum_{m}
    \gamma_m \left( k_z - k_{zn} \right)^m
,
\label{eqn-example-omegak}
\end{align}
 where $\gamma_1 = \pLightspeed + \bar{\gamma}_1$ has two contributions --
 that due to the vacuum (just $\pLightspeed$), 
 and the estimated modifications due to spatial effects ($\bar{\gamma}_1$).

{Now we arrive at the crucial point in the argment, 
 where we also decide
 to add the effects of whatever \emph{temporal} response}
 the material might have to our propagation calculation; 
 but also decide that an approximate model where we modify
 the expression above in eqn. \eqref{eqn-example-omegak}
 is sufficent.
This is the converse of the traditional procedure, 
 where the waveguide dispersion is added to whatever $\beta(\omega)$
 is appropriate for the medium's time response.
{Consequently, 
 to proceed further we only need to make very similar assumptions:}
 (a) that the two sources of dispersion are weak and so can simply be added, 
 (b) that the temporal response is assumed to have negligible effect
 on the waveguide's propagating modes, 
 and
 (c) that the temporal dispersion is
 the same at all points in the waveguide.

{The temporal response of the waveguide
 is quite naturally written in terms of 
 a wavevector dependent on a specified frequency, 
 and here is taken to have been simplified to a quadratic.
Following the recommendations of \cite{Kinsler-2009pra}, 
 I also exclude imaginary components 
 (e.g. loss or gain terms) 
 in either the wavevector reference $\beta$
 or the frequency, 
 intending to incorporate such effects as necessary at some later stage. 
Consequently, 
 for the $n$-th mode,
 we have}
~
\begin{align}
  \beta(\omega)
&=
  \beta_n
 + 
    \kappa_1 
     \left(
        \omega - \omega_n
      \right)
 + 
    \kappa_2
     \left(
        \omega - \omega_n
      \right)^2
,
\label{eqn-example-komega-temporal}
\end{align}
 where $\kappa_1 = {\pLightspeed}^{-1} + \bar{\kappa}_1$
 has two contributions --
 that due to the vacuum (just $1/\pLightspeed$), 
 and the modification due to the temporal response ($\bar{\kappa}_1$).

{In order to map this into the temporally propagated picture, 
 with a reference frequency being dependent
 on a wavevector;  
 we replace $\omega$ with $\Omega$, 
 and $\beta$ with $k_z$.
Since we have chosen a quadratic form, 
 the resulting expression 
 can then be solved\footnote{{In general -- 
 if there are more terms in the expansion,
 or $\beta$ has some specific and non-trivial analytic form --
 this step is more complicated.}}
 to find $\Omega$.}
With $\Omega_n = c k_{zn}$,
 this gives
~
\begin{align}
  \Omega(k_{z})
&=
  \Omega_n
 -
  \frac{\kappa_1}{2\kappa_2}
  \left[
    1
   \mp
    \sqrt{1 + 4 \left( k_{z} - k_{zn}\right) \kappa_2 / \kappa_1^2 }
  \right]
,
\label{eqn-example-omegak-temporal}
\end{align}
 which we could expand into a power series for small $k_{zn} - k_{z}$.
Since $\kappa_1$ includes the vacuum, 
 and dispersion $\kappa_2$ is assumed to be small, 
 we should make the top sign choice since it 
 gives the physically relevant small frequency corrections.
The linear term in the series has the coeffient $\gamma_{1T}=1/\kappa_1$
 and the quadratic term $\gamma_{2T}=\kappa_2/\kappa_1^3$.

We now combine the two dispersive effects --
 the spatial/geometric/waveguide ones based on parameters $\gamma_i$, 
 and converted temporal ones based on $\kappa_i$, 
 to give a total reference frequency based on $k_z$ of
~
\begin{align}
  \Omega'(k_z)
&=
  \Omega_n
 + 
    \bar{\gamma}_1 \left( k_{z} - k_{zn} \right)
 + 
  \sum_{m>1}
    \gamma_m \left[ \left( k_{z} - k_{zn} \right) \right]^m
,
\nonumber
\\
& \qquad
 -
  \frac{\kappa_1}{2\kappa_2}
  \left[
    1
   - 
    \sqrt{1 + 4 \left( k_{z} - k_{zn}\right) \kappa_2 / \kappa_1^2 }
  \right]
\label{eqn-example-omegak-totalised}
\\
&\simeq
  \Omega_n
 +
  \left(
    \bar{\gamma}_1 
   + 
    \frac{1}{\kappa_1} 
  \right)
  \left( k_z - k_{zn} \right)
 +
  \left(
     \gamma_2 
    - 
     \frac{\kappa_2}{\kappa_1^{3}}
  \right)
  \left( k_z - k_{zn} \right)^2
.
\label{eqn-example-omegak-totalisedp2}
\end{align}
Note that we have to avoid incorporating the vacuum contribution twice
 when summing the dispersive effects,
 {which means that $\bar{\gamma}_1$ appears instead of $\gamma_1$.}

In the complementary case, 
 useful for spatial propagation, 
 we can follow the converse -- 
 and indeed the \emph{usual} procedure -- 
 to convert
 the CW effect of the waveguide's spatial structure $\Omega(k_z)$ 
 into a frequency domain form \cite{TG-FiberOptEss,Shaw-MPOFC}.
This then can be combined with 
 temporal response of $\beta(\omega)$
 to get
~
\begin{align}
  \beta'(\omega)
&=
  \beta_0
 +
  \left( 
    \bar{\kappa}_1 
   + 
    \frac{1}{\gamma_1}
  \right)
  \left( \omega - \omega_{0} \right)
 +
  \left( 
    \kappa_2 
   - 
    \frac{\gamma_2}{\gamma_1^{3}}
  \right)
  \left( \omega - \omega_{0} \right)^2
.
\end{align}

\subsection*{Discussion}
\label{S-example-discussion}

We might have hoped to find a less approximate comparison
 between the spatially propagated and the temporally propagated 
 approaches to dispersion
 what has just been discussed.
Further, 
 although qualitative comparisons can also be made, 
 ideally we would like a specific and quantitative comparison 
 as done for e.g. 
 testing the unidirectional approximation \cite{Kinsler-2007josab}.

That this is not as simple as it might sound can be 
 demonstrated as follows:
 write two mathematically identical 
 but physically and notationally distinct propagation equations,
 i.e.
~
\begin{align}
  \partial_z {\pEfield}(\omega) 
&=
 +\imath \kappa(\omega) {\pEfield}(\omega)
\label{eqn-examples-waveZprop}
\\
  \partial_t {\pDfield}(z) 
&=
  +\imath \Omega(z) {\pDfield}(z)
\label{eqn-examples-waveTprop}
.
\end{align}
For simplicity, 
 this comparison is done in the 1+1D limit
 and for unidirectional equations; 
 the residual $\omega$ or $z$ behaviour has been 
 merged with the reference behaviour
 to give single propagation parameters $\kappa(\omega)$ or $\Omega(z)$.

Superficially, 
 the comparison between
 eqn. \eqref{eqn-examples-waveZprop} and \eqref{eqn-examples-waveTprop}
 looks promising.
However, 
 the vacuum part of $\kappa(\omega)$ is $\omega/\pLightspeed$, 
 but the vaccum part of $\Omega(z)$ is in contrast a spatial derivative, 
 i.e. $\pLightspeed\partial_z$.
Even if it were possible to match up the functional forms 
 of the non-vacuum temporal response 
 of $\kappa(\omega)-\omega/\pLightspeed$ 
 with a spatial structure giving $\Omega(z)-\pLightspeed\partial_z$, 
 the vacuum parts would remain unmatched.
Alternatively, 
 we could compare the $z$-propagated $\omega$-domain 
 to the 
 $t$-propagated $k$-domain.
However, 
 although in this case the reference behaviours match, 
 but the $k$-domain model now has a convolution
 that does not appear in the $\omega$-domain one.

As a result,
 it is unclear how well
 approximating temporal dispersion as spatial properties
 {should work}; 
 but then it is equally unclear how well 
 approximating spatial properties as temporal dispersion
 {should work either}.
Nevertheless, 
 this latter process is so ubiquitous in optics as to 
 be effectively invisible, 
 {and --
 more importantly --
 does not seem to give rise to significant problems.}
Removing either of the mismatches discussed above -- 
 either in reference behaviour or convolution --  
 requires a narrowband limit, 
 a process which necessarily obscures the parameter regime 
 where interesting tests and comparisons can (and should) be made --
 i.e. the short pulse, large bandwidth limit.
Indeed, 
 the stringent nature of the approximation would defeat
 the entire point of the comparison.

Nevertheless, 
 it may be possible to design numerical simulations 
 containing structures and temporal responses
 whose scales are carefully graduated so that 
 the mundane bandwidth issues 
 can be kept separate from the interesting propagation ones.
Such a design would, 
 as desired, 
 allow the propagation approximations (only)
 to be compared and contrasted by numerical simulation.
However, 
 this is a non-trival task,
 and one which I leave for later work.



%
\section{Pulse propagation}
\label{S-nls}

{As an example, 
 I will now consider pulse propogation in
 a simple optical fibre model \cite{Agrawal-NFO,TG-FiberOptEss,Shaw-MPOFC}, 
 where the dominant interesting effect is the 
 third order nonlinearity present silica.}
Here I compare 1+1D propagation such a waveguide,
 in order to elucidate the differences between
 the standard spatially propagated picture
 {(as discussed in \cite{Kinsler-2010pra-fchhg})}
 and the temporally propagated picture derived here.
{This comparison is made much clearer
 by not only the use of the factorization method in both cases, 
 but also because it enabled
 both derivations to closely follow the same steps and approximations.}

The assumptions made are those of 
 transverse fields, 
 weak dispersive corrections, 
 and weakly nonlinear response; 
 these all allow us to decouple
 the forward and backward wave equations.
This decoupling {means that 
 without using any extra approximations,
 we can
 simplify our description and treat} forward only pulse propagation.
The specific example chosen here
 is for an instantaneous cubic nonlinearity, 
 but it is easily generalized to non-instantaneous cases
 or even other scalar nonlinearities.

%
\subsection{Spatially propagated NLSE}

{The commonly used
 spatially propagated NLS for ${E}_x^{+}(\omega)$
 \cite{Boyd-NLO,Brabec-K-1997prl,Laegsgaard-2007oe}}
 based on a {directional factorization} 
 \cite{Kinsler-2010pra-fchhg}
 can be written
~
\begin{align}
  \partial_z {\pEfield}_x^{+}
&=
 +
  \imath 
    \beta
  {\pEfield}_x^{+}
 ~
 +
  \sum_m
    \kappa_m . 
      \left( \omega - \omega_0 \right)^m
    {\pEfield}_x^{+}
\nonumber
\\
& \qquad\qquad\quad
 ~
 +
  \frac{\imath k_0^2}
       {2 \beta}
  \mathscr{F}
  \left[
    \chi^{(3)}
    {\pEfield}_x^2(t)
    {\pEfield}_x^{+}(t)
  \right]
,
\label{eqn-eg-Enls}
\end{align}
 where $\mathscr{F}[...]$ is the Fourier transform that 
 converts the nonlinear polarization term
 into its frequency domain form.
Here we have chosen a fixed, 
 non-dispersive reference wavevector $\beta$, 
 the temporal response of the propagation medium 
 is encoded in the coefficients $\kappa_j$.
For the temporal (time response) of the material, 
 this is an in-principle exact re-representation of that response
 as a Taylor series expansion; 
 although in practise the series is usually truncated after a few terms
 and the propagating fields restricted to within a bandwidth where 
 only those terms are significant.

The important point to remember is that
 in this spatially propagated picture we do know the full time history, 
 at least in a computational sense --
 i.e. as we compute a solution to the propagation equation.
This ``computational past'' \cite{Kinsler-2014arXiv-negfreq,Ajaib-2013arxiv},
 comprising all previously computed values of the field properties
 and material state means that a full frequency spectrum and response
 is known.
As a result,
 the time response of the material \emph{can} be directly encoded in
 an accurate and useful dispersion relation, 
 and then be approximated as a Taylor series in $\omega$.
A depiction of this spatial propagation scheme
 was shown in fig. \ref{fig-reflect-z}

Since this propagation is assumed to be taking place
 in a waveguide, 
 we also have its geometric --
 wavelength sensitive --
 properties to consider.
However, 
 as discussed in the previous section we typically just re-represent those
 \emph{as if they were instead generated by a temporal response}
 (see e.g. \cite{TG-FiberOptEss,Shaw-MPOFC})
Thus the coefficients $\kappa_j$
 usually 
 combine both the effect of time response \emph{and} geometric properties
 into a single total dispersion; 
 this is the very basis of dispersion compensation 
 in optical fibres, 
 since the material (temporal) dispersion 
 is offset by the waveguide design (spatial dispersion)
 \cite{TG-FiberOptEss}.

%
\subsection{Temporally propagated NLSE}

Using the approach derived and presented in this paper, 
 we can develop a time propagated counterpart to eqn. \eqref{eqn-eg-Enls}.
In a purely mathematical sense, 
 it is, 
 apart from notation, 
 identical to eqn. \eqref{eqn-eg-Enls}.
However, 
 in physical terms,
 the change is not trivial, 
 since the physical meaning and boundary conditions differ significantly.
Starting from eqn. \eqref{eqn-uni-dtD}
 we can get a propagation equation
 in wavevector space for ${\pDfield}_x^{+}(k)$ which reads 
~
\begin{align}
  \partial_t {\pDfield}_x^{+}
&=
 +
  \imath 
    \Omega
  {\pDfield}_x^{+}
 ~
 +
  \sum_m
    \gamma_m . 
    \left( k_z - k_{z0} \right)^m
    {\pDfield}_x^{+}
\nonumber
\\
& \qquad\qquad\quad
 ~
 +
  \frac{\imath \Omega}
       {2 a_{\pPermittivity}}
  \mathscr{G}
  \left[
    \chi^{(3)}
    {\pDfield}_x^2(z)
    {\pDfield}_x^{+}(z)
  \right]
\label{eqn-eg-Dnls}
,
\end{align}
 where $\mathscr{G}[...]$ is the Fourier transform that 
 converts the spatially localized nonlinear polarization 
 into its wavevector domain form.
\emph{Here,}
 all dispersive behaviour has been combined into the coefficients $\gamma_j$, 
 which mimic the common notion of spatial dispersion.
Such terms can arise directly from a Taylor expansion in $k$ 
 of a relevant CW dispersion relation 
 about some suitable reference wavevector;
 however, 
 although we can use the expansion to represent the spatial structure
 of the propagation medium (e.g. a waveguide), 
 the conversion from spatial structure to $k$ expansion
 is only relatively straightforward in the CW limit.

If the propagation medium also has a temporal response
 (i.e. temporal dispersion),
 and if we do not want to model it explicitly, 
 then we can use the approach described in the previous section, 
 where the temporal response of the material
 was (approximately) represented 
 as if it were instead due to spatial properties.

Finally, 
 note that in ordinary temporally propagated simulations, 
 the computational past and the physical (temporal) past 
 are identical; 
 this is not the case for spatially propagated simulations, 
 as mentioned above.
A depiction of the temporal propagation scheme
 was shown in fig. \ref{fig-reflect-t}

%
\section{Conclusion}
\label{S-conclude}

I have derived a general first order wave equation
 for uni-directional pulse propagation in time.
{The scope was kept as general as possible, 
 allowing for
 arbitrary dielectric polarization, 
 magnetization,}
 diffraction, 
 and free electric charge and currents.
This propagation equation has a utility all of its own, 
 allowing efficient unidirectional propagation
 in a strictly causal manner 
 (i.e. time directed).
However,
 it requires comparable approximations to those used  
 in deriving spatially propagated unidirectional propagation equations, 
 although the specific tradeoffs are different.

The propagation equation was derived 
 by first factorizing the second order wave equation for ${\pDfieldv}$
 into an exact bi-directional model.
{Next, 
 I applied
 the same type of well defined type of approximation
 to all non-trivial effects (e.g. nonlinearity, diffraction), 
 and so reduced the bi-directional propagation equations
 down to a simpler} first order uni-directional wave equation.
One feature of this factorization method
 is that it provides {an easy way to do a term-to-term comparison
 of the exact bi-directional theory
 and approximate uni-directional descriptions.}

In addition to its use where the causal simulation of a propagation problem
 is of strict concern, 
 it also illuminates the process of handling 
 material response and spatial structure in waveguides
 by constructing a single combined dispersion model.
As discussed in section \ref{S-dispersion}, 
 we can now see that this simple addition of 
 temporal and spatial dispersions in an ad hoc manner
 stands on rather poor foundations,
 particularly in the few-cycle pulse limit.
Further, 
 it is hard to see any straightforward way of testing the limitations 
 of this common procedure, 
 which poses an interesting challenge for the future.
That said, 
 in most applications, 
 we expect that a combined dispersion will remain
 a perfectly adequate approximation, 
 since the pulses are rarely so short, 
 and it seems reasonable to assume 
 that the character of the pulse evolution should remain similar, 
 even if details might differ.
Lastly, 
 the lessons learnt from this investigation could also be applied to other
 types of wave propagation, 
 most notably acoustic systems (see e.g. \cite{Kinsler-2012arXiv-fbacou}).

%

\begin{acknowledgments}

The great majority of the work here was done whilst at
 Imperial College London, 
 supported by EPSRC (grant number EP/K003305/1); 
 but finished when at 
 Lancaster University, 
 again supported by EPSRC (the Alpha-X project EP/N028694/1).


\end{acknowledgments}

%


\begin{thebibliography}{42}
\expandafter\ifx\csname natexlab\endcsname\relax\def\natexlab#1{#1}\fi
\expandafter\ifx\csname bibnamefont\endcsname\relax
  \def\bibnamefont#1{#1}\fi
\expandafter\ifx\csname bibfnamefont\endcsname\relax
  \def\bibfnamefont#1{#1}\fi
\expandafter\ifx\csname citenamefont\endcsname\relax
  \def\citenamefont#1{#1}\fi
\expandafter\ifx\csname url\endcsname\relax
  \def\url#1{\texttt{#1}}\fi
\expandafter\ifx\csname urlprefix\endcsname\relax\def\urlprefix{URL }\fi
\providecommand{\bibinfo}[2]{#2}
\providecommand{\eprint}[2][]{\url{#2}}

\bibitem[{\citenamefont{Brabec and Krausz}(2000)}]{Brabec-K-2000rmp}
\bibinfo{author}{\bibfnamefont{T.}~\bibnamefont{Brabec}} \bibnamefont{and}
  \bibinfo{author}{\bibfnamefont{F.}~\bibnamefont{Krausz}},
  \\ \bibinfo{journal}{Rev. Mod. Phys.} \textbf{\bibinfo{volume}{72}},
  \bibinfo{pages}{545} (\bibinfo{year}{2000}),
  \\ \XDOI{10.1103/RevModPhys.72.545}.

\bibitem[{\citenamefont{Corkum}(1993)}]{Corkum-1993prl}
\bibinfo{author}{\bibfnamefont{P.~B.} \bibnamefont{Corkum}},
  \\ \bibinfo{journal}{Phys. Rev. Lett.} \textbf{\bibinfo{volume}{71}},
  \bibinfo{pages}{1994 } (\bibinfo{year}{1993}),
  \\ \XDOI{10.1103/PhysRevLett.71.1994}.

\bibitem[{\citenamefont{Schafer et~al.}(1993)\citenamefont{Schafer, Yang,
  DiMauro, and Kulander}}]{Schafer-YDK-1993prl}
\bibinfo{author}{\bibfnamefont{K.~J.} \bibnamefont{Schafer}},
  \bibinfo{author}{\bibfnamefont{B.}~\bibnamefont{Yang}},
  \bibinfo{author}{\bibfnamefont{L.~F.} \bibnamefont{DiMauro}},
  \bibnamefont{and} \bibinfo{author}{\bibfnamefont{K.~C.}
  \bibnamefont{Kulander}}, \\ \bibinfo{journal}{Phys. Rev. Lett.}
  \textbf{\bibinfo{volume}{70}}, \bibinfo{pages}{1599 } (\bibinfo{year}{1993}),
  \\ \XDOI{10.1103/PhysRevLett.70.1599}.

\bibitem[{\citenamefont{Fuji et~al.}(2005)\citenamefont{Fuji, Rauschenberger,
  Gohle, Apolonski, Udem, Yakovlev, Tempea, Hansch, and
  Krausz}}]{Fuji-RGAUYTHK-2005njp}
\bibinfo{author}{\bibfnamefont{T.}~\bibnamefont{Fuji}},
  \bibinfo{author}{\bibfnamefont{J.}~\bibnamefont{Rauschenberger}},
  \bibinfo{author}{\bibfnamefont{C.}~\bibnamefont{Gohle}},
  \bibinfo{author}{\bibfnamefont{A.}~\bibnamefont{Apolonski}},
  \bibinfo{author}{\bibfnamefont{T.}~\bibnamefont{Udem}},
  \bibinfo{author}{\bibfnamefont{V.~S.} \bibnamefont{Yakovlev}},
  \bibinfo{author}{\bibfnamefont{G.}~\bibnamefont{Tempea}},
  \bibinfo{author}{\bibfnamefont{T.~W.} \bibnamefont{Hansch}},
  \bibnamefont{and} \bibinfo{author}{\bibfnamefont{F.}~\bibnamefont{Krausz}},
  \\ \bibinfo{journal}{New J. Phys.} \textbf{\bibinfo{volume}{7}},
  \bibinfo{pages}{116} (\bibinfo{year}{2005}),
  \\ \XDOI{10.1088/1367-2630/7/1/116}.

\bibitem[{\citenamefont{Radnor et~al.}(2008)\citenamefont{Radnor, Chipperfield,
  Kinsler, and New}}]{Radnor-CKN-2008pra}
\bibinfo{author}{\bibfnamefont{S.~B.~P.} \bibnamefont{Radnor}},
  \bibinfo{author}{\bibfnamefont{L.~E.} \bibnamefont{Chipperfield}},
  \bibinfo{author}{\bibfnamefont{P.}~\bibnamefont{Kinsler}}, 
  \bibinfo{author}{\bibfnamefont{G.~H.~C.} \bibnamefont{New}},
  \\ \bibinfo{journal}{Phys. Rev. A} \textbf{\bibinfo{volume}{77}},
  \bibinfo{pages}{033806} (\bibinfo{year}{2008}), \\ \XARXIV{0803.3597},
  \\ \XDOI{10.1103/PhysRevA.77.033806}.

\bibitem[{\citenamefont{Solanp{\"a}{\"a}
  et~al.}(2014)\citenamefont{Solanp{\"a}{\"a}, Budagosky, Shvetsov-Shilovski,
  Castro, Rubio, and R{\"a}s{\"a}nen}}]{Solanpaa-BSCRR-2014pra}
\bibinfo{author}{\bibfnamefont{J.}~\bibnamefont{Solanp{\"a}{\"a}}},
  \bibinfo{author}{\bibfnamefont{J.~A.} \bibnamefont{Budagosky}},
  \bibinfo{author}{\bibfnamefont{N.~I.} \bibnamefont{Shvetsov-Shilovski}},
  \bibinfo{author}{\bibfnamefont{A.}~\bibnamefont{Castro}},
  \bibinfo{author}{\bibfnamefont{A.}~\bibnamefont{Rubio}}, \bibnamefont{and}
  \bibinfo{author}{\bibfnamefont{E.}~\bibnamefont{R{\"a}s{\"a}nen}},
  \\ \bibinfo{journal}{Phys. Rev. A} \textbf{\bibinfo{volume}{90}},
  \bibinfo{pages}{053402} (\bibinfo{year}{2014}),
  \\ \XDOI{10.1103/PhysRevA.90.053402}.

\bibitem[{\citenamefont{Alfano and
  Shapiro}(1970{\natexlab{a}})}]{Alfano-S-1970prl}
\bibinfo{author}{\bibfnamefont{R.~R.} \bibnamefont{Alfano}} \bibnamefont{and}
  \bibinfo{author}{\bibfnamefont{S.~L.} \bibnamefont{Shapiro}},
  \\ \bibinfo{journal}{Phys. Rev. Lett.} \textbf{\bibinfo{volume}{24}},
  \bibinfo{pages}{584} (\bibinfo{year}{1970}{\natexlab{a}}),
  \\ \XDOI{10.1103/PhysRevLett.24.584}.

\bibitem[{\citenamefont{Alfano and
  Shapiro}(1970{\natexlab{b}})}]{Alfano-S-1970prl-2}
\bibinfo{author}{\bibfnamefont{R.~R.} \bibnamefont{Alfano}} \bibnamefont{and}
  \bibinfo{author}{\bibfnamefont{S.~L.} \bibnamefont{Shapiro}},
  \\ \bibinfo{journal}{Phys. Rev. Lett.} \textbf{\bibinfo{volume}{24}},
  \bibinfo{pages}{592} (\bibinfo{year}{1970}{\natexlab{b}}),
  \\ \XDOI{10.1103/PhysRevLett.24.592}.

\bibitem[{\citenamefont{Dudley et~al.}(2006)\citenamefont{Dudley, Genty, and
  Coen}}]{Dudley-GC-2006rmp}
\bibinfo{author}{\bibfnamefont{J.~M.} \bibnamefont{Dudley}},
  \bibinfo{author}{\bibfnamefont{G.}~\bibnamefont{Genty}}, \bibnamefont{and}
  \bibinfo{author}{\bibfnamefont{S.}~\bibnamefont{Coen}},
  \\ \bibinfo{journal}{Rev. Mod. Phys.} \textbf{\bibinfo{volume}{78}},
  \bibinfo{pages}{1135 } (\bibinfo{year}{2006}),
  \\ \XDOI{10.1103/RevModPhys.78.1135}.

\bibitem[{\citenamefont{Solli et~al.}(2007)\citenamefont{Solli, Ropers,
  Koonath, and Jalali}}]{Solli-RKJ-2007n}
\bibinfo{author}{\bibfnamefont{D.~R.} \bibnamefont{Solli}},
  \bibinfo{author}{\bibfnamefont{C.}~\bibnamefont{Ropers}},
  \bibinfo{author}{\bibfnamefont{P.}~\bibnamefont{Koonath}}, \bibnamefont{and}
  \bibinfo{author}{\bibfnamefont{B.}~\bibnamefont{Jalali}},
  \\ \bibinfo{journal}{Nature} \textbf{\bibinfo{volume}{450}},
  \bibinfo{pages}{1054} (\bibinfo{year}{2007}),
  \\ \XDOI{10.1038/nature06402}.

\bibitem[{\citenamefont{Braun et~al.}(1995)\citenamefont{Braun, Korn, Liu, Du,
  Squier, and Mourou}}]{Braun-KLDSM-1995ol}
\bibinfo{author}{\bibfnamefont{A.}~\bibnamefont{Braun}},
  \bibinfo{author}{\bibfnamefont{G.}~\bibnamefont{Korn}},
  \bibinfo{author}{\bibfnamefont{X.}~\bibnamefont{Liu}},
  \bibinfo{author}{\bibfnamefont{D.}~\bibnamefont{Du}},
  \bibinfo{author}{\bibfnamefont{J.}~\bibnamefont{Squier}}, \bibnamefont{and}
  \bibinfo{author}{\bibfnamefont{G.}~\bibnamefont{Mourou}},
  \\ \bibinfo{journal}{Opt. Lett.} \textbf{\bibinfo{volume}{20}},
  \bibinfo{pages}{73} (\bibinfo{year}{1995}),
  \\ \XDOI{10.1364/OL.20.000073}.

\bibitem[{\citenamefont{Chin et~al.}(1999)\citenamefont{Chin, Petit, Borne, and
  Miyazaki}}]{Chin-PBM-1999jjapl}
\bibinfo{author}{\bibfnamefont{S.~L.} \bibnamefont{Chin}},
  \bibinfo{author}{\bibfnamefont{S.}~\bibnamefont{Petit}},
  \bibinfo{author}{\bibfnamefont{F.}~\bibnamefont{Borne}}, \bibnamefont{and}
  \bibinfo{author}{\bibfnamefont{K.}~\bibnamefont{Miyazaki}},
  \\ \bibinfo{journal}{Jpn. J. Appl. Phys.} \textbf{\bibinfo{volume}{38}},
  \bibinfo{pages}{L126} (\bibinfo{year}{1999}),
  \\ \XDOI{10.1143/JJAP.38.L126}.

\bibitem[{\citenamefont{Milchberg et~al.}(2014)\citenamefont{Milchberg, Chen,
  Cheng, Jhajj, Palastro, Rosenthal, Varma, Wahlstrand, and
  Zahedpour}}]{Milchberg-CCJPRVWZ-2014pp}
\bibinfo{author}{\bibfnamefont{H.~M.} \bibnamefont{Milchberg}},
  \bibinfo{author}{\bibfnamefont{Y.~H.} \bibnamefont{Chen}},
  \bibinfo{author}{\bibfnamefont{Y.~H.} \bibnamefont{Cheng}},
  \bibinfo{author}{\bibfnamefont{N.}~\bibnamefont{Jhajj}},
  \bibinfo{author}{\bibfnamefont{J.~P.} \bibnamefont{Palastro}},
  \bibinfo{author}{\bibfnamefont{E.~W.} \bibnamefont{Rosenthal}},
  \bibinfo{author}{\bibfnamefont{S.}~\bibnamefont{Varma}},
  \bibinfo{author}{\bibfnamefont{J.~K.} \bibnamefont{Wahlstrand}},
  \bibinfo{author}{\bibfnamefont{S.}~\bibnamefont{Zahedpour}},
  \\ \bibinfo{journal}{Phys. Plasmas} \textbf{\bibinfo{volume}{21}},
  \bibinfo{pages}{100901} (\bibinfo{year}{2014}),
  \\ \XDOI{10.1063/1.4896722}.

\bibitem[{\citenamefont{Kinsler}(2010)}]{Kinsler-2010pra-fchhg}
\bibinfo{author}{\bibfnamefont{P.}~\bibnamefont{Kinsler}},
  \\ \bibinfo{journal}{Phys. Rev. A} \textbf{\bibinfo{volume}{81}},
  \bibinfo{pages}{013819} (\bibinfo{year}{2010}), \\ \XARXIV{0810.5689},
  \\ \XDOI{10.1103/PhysRevA.81.013819}.

\bibitem[{\citenamefont{Kinsler}(2012{\natexlab{a}})}]{Kinsler-2012arxiv-fbrad}
\bibinfo{author}{\bibfnamefont{P.}~\bibnamefont{Kinsler}}
  (\bibinfo{year}{2012}{\natexlab{a}}), \\ \XARXIV{1210.6794}.

\bibitem[{\citenamefont{Kinsler}(2014)}]{Kinsler-2014arXiv-negfreq}
\bibinfo{author}{\bibfnamefont{P.}~\bibnamefont{Kinsler}}
  (\bibinfo{year}{2014}), \\ \bibinfo{note}{``Negative Frequency Waves? Or: What I
  talk about when I talk about propagation''}, \\ \XARXIV{1408.0128}.

\bibitem[{\citenamefont{Kinsler}(2011)}]{Kinsler-2011ejp}
\bibinfo{author}{\bibfnamefont{P.}~\bibnamefont{Kinsler}},
  \\ \bibinfo{journal}{Eur. J. Phys.} \textbf{\bibinfo{volume}{32}},
  \bibinfo{pages}{1687} (\bibinfo{year}{2011}), \\ \bibinfo{note}{the arXiv
  version has additional appendices}, \\ \XARXIV{1106.1792},
  \\ \XDOI{10.1088/0143-0807/32/6/022}.

\bibitem[{\citenamefont{Yee}(1966)}]{Yee-1966tap}
\bibinfo{author}{\bibfnamefont{K.~S.} \bibnamefont{Yee}},
  \\ \bibinfo{journal}{IEEE Trans. Antennas Propagat.}
  \textbf{\bibinfo{volume}{14}}, \bibinfo{pages}{302} (\bibinfo{year}{1966}),
  \\ \XDOI{10.1109/TAP.1966.1138693}.

\bibitem[{\citenamefont{Kinsler}(2012{\natexlab{b}})}]{Kinsler-2012arXiv-fbacou}
\bibinfo{author}{\bibfnamefont{P.}~\bibnamefont{Kinsler}}
  (\bibinfo{year}{2012}{\natexlab{b}}), \\ \bibinfo{note}{``Acoustic waves: should
  they be propagated forward in time, or forward in space?''},
  \\ \XARXIV{1202.0714}.

\bibitem[{\citenamefont{Agrawal}(2007)}]{Agrawal-NFO}
\bibinfo{author}{\bibfnamefont{G.~P.} \bibnamefont{Agrawal}},\\
  \emph{\bibinfo{title}{Nonlinear Fiber Optics}} \\
 (\bibinfo{publisher}{Academic
  Press Inc.}, \bibinfo{address}{Boston}, \bibinfo{year}{2007}),
  \bibinfo{edition}{4th} ed., \\
  ISBN \bibinfo{isbn}{978-0-12-369516-1}.

\bibitem[{\citenamefont{Berge and Skupin}(2009)}]{Berge-S-2009dcdsa}
\bibinfo{author}{\bibfnamefont{L.}~\bibnamefont{Berge}} \bibnamefont{and}
  \bibinfo{author}{\bibfnamefont{S.}~\bibnamefont{Skupin}},
  \\ \bibinfo{journal}{Discrete Continuous Dyn. Syst. (DCDS-A)}
  \textbf{\bibinfo{volume}{23}}, \bibinfo{pages}{1099 } (\bibinfo{year}{2009}),
  \\ \XDOI{10.3934/dcds.2009.23.1099}.

\bibitem[{\citenamefont{Ciraci et~al.}(2013)\citenamefont{Ciraci, Pendry, and
  Smith}}]{Ciraci-PS-2013cphc}
\bibinfo{author}{\bibfnamefont{C.}~\bibnamefont{Ciraci}},
  \bibinfo{author}{\bibfnamefont{J.~B.} \bibnamefont{Pendry}},
  \bibnamefont{and} \bibinfo{author}{\bibfnamefont{D.~R.} \bibnamefont{Smith}},
  \\ \bibinfo{journal}{ChemPhysChem} \textbf{\bibinfo{volume}{14}},
  \bibinfo{pages}{1109} (\bibinfo{year}{2013}),
  \\ \XDOI{10.1002/cphc.201200992}.

\bibitem[{\citenamefont{Oskooi et~al.}(2010)\citenamefont{Oskooi, Roundy,
  Ibanescu, Bermel, Joannopoulos, and Johnson}}]{Oskooi-RIBJJ-2010cpc}
\bibinfo{author}{\bibfnamefont{A.~F.} \bibnamefont{Oskooi}},
  \bibinfo{author}{\bibfnamefont{D.}~\bibnamefont{Roundy}},
  \bibinfo{author}{\bibfnamefont{M.}~\bibnamefont{Ibanescu}},
  \bibinfo{author}{\bibfnamefont{P.}~\bibnamefont{Bermel}},
  \bibinfo{author}{\bibfnamefont{J.~D.} \bibnamefont{Joannopoulos}},
  \bibnamefont{and} \bibinfo{author}{\bibfnamefont{S.~G.}
  \bibnamefont{Johnson}}, \\ \bibinfo{journal}{Comput. Phys. Commun.}
  \textbf{\bibinfo{volume}{181}}, \bibinfo{pages}{687} (\bibinfo{year}{2010}),
  \\ \XDOI{10.1016/j.cpc.2009.11.008}.

\bibitem[{\citenamefont{Carter}(1995)}]{Carter-1995pra}
\bibinfo{author}{\bibfnamefont{S.~J.} \bibnamefont{Carter}},
  \\ \bibinfo{journal}{Phys. Rev. A} \textbf{\bibinfo{volume}{51}},
  \bibinfo{pages}{3274 } (\bibinfo{year}{1995}),
  \\ \XDOI{10.1103/PhysRevA.51.3274}.

\bibitem[{\citenamefont{Ferrando et~al.}(2005)\citenamefont{Ferrando, Zacares,
  de~Cordoba, Binosi, and Montero}}]{Ferrando-ZCBM-2005pre}
\bibinfo{author}{\bibfnamefont{A.}~\bibnamefont{Ferrando}},
  \bibinfo{author}{\bibfnamefont{M.}~\bibnamefont{Zacares}},
  \bibinfo{author}{\bibfnamefont{P.~F.} \bibnamefont{de~Cordoba}},
  \bibinfo{author}{\bibfnamefont{D.}~\bibnamefont{Binosi}}, \bibnamefont{and}
  \bibinfo{author}{\bibfnamefont{A.}~\bibnamefont{Montero}},
  \\ \bibinfo{journal}{Phys. Rev. E} \textbf{\bibinfo{volume}{71}},
  \bibinfo{pages}{016601} (\bibinfo{year}{2005}),
  \\ \XDOI{10.1103/PhysRevE.71.016601}.

\bibitem[{\citenamefont{Blow and Wood}(1989)}]{Blow-W-1989jqe}
\bibinfo{author}{\bibfnamefont{K.~J.} \bibnamefont{Blow}} \bibnamefont{and}
  \bibinfo{author}{\bibfnamefont{D.}~\bibnamefont{Wood}},
  \\ \bibinfo{journal}{IEEE J. Quantum Electronics} \textbf{\bibinfo{volume}{25}},
  \bibinfo{pages}{2665} (\bibinfo{year}{1989}),
  \\ \XDOI{10.1109/3.40655}.

\bibitem[{\citenamefont{Kinsler}(2009)}]{Kinsler-2009pra}
\bibinfo{author}{\bibfnamefont{P.}~\bibnamefont{Kinsler}},
  \\ \bibinfo{journal}{Phys. Rev. A} \textbf{\bibinfo{volume}{79}},
  \bibinfo{pages}{023839} (\bibinfo{year}{2009}), 
  \\ \XARXIV{0901.2466},
  \\ \XDOI{10.1103/PhysRevA.79.023839}.

\bibitem[{\citenamefont{Kinsler et~al.}(2005)\citenamefont{Kinsler, Radnor, and
  New}}]{Kinsler-RN-2005pra}
\bibinfo{author}{\bibfnamefont{P.}~\bibnamefont{Kinsler}},
  \bibinfo{author}{\bibfnamefont{S.~B.~P.} \bibnamefont{Radnor}},
  \bibnamefont{and} \bibinfo{author}{\bibfnamefont{G.~H.~C.}
  \bibnamefont{New}}, \\ \bibinfo{journal}{Phys. Rev. A}
  \textbf{\bibinfo{volume}{72}}, \bibinfo{pages}{063807}
  (\bibinfo{year}{2005}), \\ \XARXIV{physics/0611215},
  \\ \XDOI{10.1103/PhysRevE.75.066603}.

\bibitem[{\citenamefont{Kinsler et~al.}(2006)\citenamefont{Kinsler, New, and
  Tyrrell}}]{Kinsler-NT-2006-phnlo}
\bibinfo{author}{\bibfnamefont{P.}~\bibnamefont{Kinsler}},
  \bibinfo{author}{\bibfnamefont{G.~H.~C.} \bibnamefont{New}},
  \bibnamefont{and} \bibinfo{author}{\bibfnamefont{J.~C.~A.}
  \bibnamefont{Tyrrell}} (\bibinfo{year}{2006}), \\ \XARXIV{physics/0611213}.

\bibitem[{\citenamefont{Boyd}(2008)}]{Boyd-NLO}
\bibinfo{author}{\bibfnamefont{R.~W.} \bibnamefont{Boyd}},\\
  \emph{\bibinfo{title}{Nonlinear Optics}} \\
 (\bibinfo{publisher}{Academic Press Inc.}, \bibinfo{address}{New York}, \bibinfo{year}{2008}),
  \bibinfo{edition}{3rd} ed., \\
 ISBN \bibinfo{isbn}{978-0-12-369470-6}.

\bibitem[{\citenamefont{Harris}(1966)}]{Harris-1966apl}
\bibinfo{author}{\bibfnamefont{S.~E.} \bibnamefont{Harris}},
  \\ \bibinfo{journal}{Appl. Phys. Lett.} \textbf{\bibinfo{volume}{9}},
  \bibinfo{pages}{114} (\bibinfo{year}{1966}),
  \\ \XDOI{10.1063/1.1754668}.

\bibitem[{\citenamefont{Kinsler}(2007)}]{Kinsler-2007josab}
\bibinfo{author}{\bibfnamefont{P.}~\bibnamefont{Kinsler}}, \\ \bibinfo{journal}{J.
  Opt. Soc. Am. B} \textbf{\bibinfo{volume}{24}}, \bibinfo{pages}{2363}
  (\bibinfo{year}{2007}), \\ \XARXIV{0707.0986},
  \\ \XDOI{10.1364/JOSAB.24.002363}.

\bibitem[{\citenamefont{Brabec and Krausz}(1997)}]{Brabec-K-1997prl}
\bibinfo{author}{\bibfnamefont{T.}~\bibnamefont{Brabec}} \bibnamefont{and}
  \bibinfo{author}{\bibfnamefont{F.}~\bibnamefont{Krausz}},
  \\ \bibinfo{journal}{Phys. Rev. Lett.} \textbf{\bibinfo{volume}{78}},
  \bibinfo{pages}{3282} (\bibinfo{year}{1997}),
  \\ \XDOI{10.1103/PhysRevLett.78.3282}.

\bibitem[{\citenamefont{Kinsler and New}(2003)}]{Kinsler-N-2003pra}
\bibinfo{author}{\bibfnamefont{P.}~\bibnamefont{Kinsler}} \bibnamefont{and}
  \bibinfo{author}{\bibfnamefont{G.~H.~C.} \bibnamefont{New}},
  \\ \bibinfo{journal}{Phys. Rev. A} \textbf{\bibinfo{volume}{67}},
  \bibinfo{pages}{023813} (\bibinfo{year}{2003}), \\ \XARXIV{physics/0212016},
  \\ \XDOI{10.1103/PhysRevA.67.023813}.

\bibitem[{\citenamefont{Geissler et~al.}(1999)\citenamefont{Geissler, Tempea,
  Scrinzi, Schn{\"u}rer, Krausz, and Brabec}}]{Geissler-TSSKB-1999prl}
\bibinfo{author}{\bibfnamefont{M.}~\bibnamefont{Geissler}},
  \bibinfo{author}{\bibfnamefont{G.}~\bibnamefont{Tempea}},
  \bibinfo{author}{\bibfnamefont{A.}~\bibnamefont{Scrinzi}},
  \bibinfo{author}{\bibfnamefont{M.}~\bibnamefont{Schn{\"u}rer}},
  \bibinfo{author}{\bibfnamefont{F.}~\bibnamefont{Krausz}}, \bibnamefont{and}
  \bibinfo{author}{\bibfnamefont{T.}~\bibnamefont{Brabec}},
  \\ \bibinfo{journal}{Phys. Rev. Lett.} \textbf{\bibinfo{volume}{83}},
  \bibinfo{pages}{2930} (\bibinfo{year}{1999}),
  \\ \XDOI{10.1103/PhysRevLett.83.2930}.

\bibitem[{\citenamefont{Scalora and Crenshaw}(1994)}]{Scalora-C-1994oc}
\bibinfo{author}{\bibfnamefont{M.}~\bibnamefont{Scalora}} \bibnamefont{and}
  \bibinfo{author}{\bibfnamefont{M.~E.} \bibnamefont{Crenshaw}},
  \\ \bibinfo{journal}{Opt. Comm.} \textbf{\bibinfo{volume}{108}},
  \bibinfo{pages}{191} (\bibinfo{year}{1994}),
  \\ \XDOI{10.1016/0030-4018(94)90647-5}.

\bibitem[{\citenamefont{Kinsler and New}(2004)}]{Kinsler-N-2004pra}
\bibinfo{author}{\bibfnamefont{P.}~\bibnamefont{Kinsler}} \bibnamefont{and}
  \bibinfo{author}{\bibfnamefont{G.~H.~C.} \bibnamefont{New}},
  \\ \bibinfo{journal}{Phys. Rev. A} \textbf{\bibinfo{volume}{69}},
  \bibinfo{pages}{013805} (\bibinfo{year}{2004}), \\ \XARXIV{physics/0411120v1},
  \\ \XDOI{10.1103/PhysRevA.69.013805}.

\bibitem[{\citenamefont{Kinsler and New}(2005)}]{Kinsler-N-2005pra}
\bibinfo{author}{\bibfnamefont{P.}~\bibnamefont{Kinsler}} \bibnamefont{and}
  \bibinfo{author}{\bibfnamefont{G.~H.~C.} \bibnamefont{New}},
  \\ \bibinfo{journal}{Phys. Rev. A} \textbf{\bibinfo{volume}{72}},
  \bibinfo{pages}{033804} (\bibinfo{year}{2005}), \\ \XARXIV{physics/0606111},
  \\ \XDOI{10.1103/PhysRevA.72.033804}.

\bibitem[{\citenamefont{Thyagarajan and Ghatak}(2007)}]{TG-FiberOptEss}
\bibinfo{author}{\bibfnamefont{K.~S.} \bibnamefont{Thyagarajan}}
  \bibnamefont{and} \bibinfo{author}{\bibfnamefont{A.}~\bibnamefont{Ghatak}},\\
  \emph{\bibinfo{title}{Fiber Optic Essentials}}\\
  (\bibinfo{publisher}{Wiley-IEEE Press}, \bibinfo{year}{2007}), \\
  ISBN
  \bibinfo{isbn}{978-0-470-09742-7},
  \\ \XDOI{10.1002/9780470152560}.

\bibitem[{\citenamefont{Shaw}(2004)}]{Shaw-MPOFC}
\bibinfo{author}{\bibfnamefont{J.~K.} \bibnamefont{Shaw}},\\
  \emph{\bibinfo{title}{Mathematical Principles of Optical Fiber
  Communications}}, \\
  CBMS-NSF Regional Conference Series in Applied Mathematics\\
  (\bibinfo{publisher}{SIAM}, \bibinfo{year}{2004}), \\
 ISBN
  \bibinfo{isbn}{978-0-89871-708-2},
  \\ \XDOI{10.1137/1.9780898717082}.

\bibitem[{\citenamefont{Laegsgaard}(2007)}]{Laegsgaard-2007oe}
\bibinfo{author}{\bibfnamefont{J.}~\bibnamefont{Laegsgaard}},
  \\ \bibinfo{journal}{Opt. Express} \textbf{\bibinfo{volume}{15}},
  \bibinfo{pages}{16110} (\bibinfo{year}{2007}),
  \\ \XDOI{10.1364/OE.15.016110}.

\bibitem[{\citenamefont{Ajaib}(2013)}]{Ajaib-2013arxiv}
\bibinfo{author}{\bibfnamefont{M.~A.} \bibnamefont{Ajaib}}
  (\bibinfo{year}{2013}), \\ \XARXIV{1302.5601}.

\end{thebibliography}

%
\renewcommand{\theequation}{\Alph{section}.\arabic{subsection}.\arabic{equation}}

\appendix

%
\section{Factorizing}
\label{S-factorize}

Here I present a simple overview of the mathematics
 of the factorization procedure, 
 since full details can be found in \cite{Ferrando-ZCBM-2005pre}.
In the calculations below, 
 I alter the physics but not the mathematical process 
 (of \cite{Ferrando-ZCBM-2005pre})
 to instead transform into frequency space, 
 where the $t$-derivative $\partial_t$ is converted to $\imath \omega$.
This presentation closely matches that by Kinsler \cite{Kinsler-2010pra-fchhg},
 but altered away from the spatial propagation scheme used there
 so as to be appropriate for temporal propagation used here.
Also, 
 we have that $\Omega^2 = {\pLightspeed}^2 k^2 / n^2$,
 and the unspecified residual term is denoted $Q$.
The second order wave equation can then be written
~
\begin{align}
  \left[
    \partial_t^2 
   +
    \Omega^2
  \right]
  D
&=
 -Q
\\
  \left[
   -\omega^2 + \Omega^2
  \right]
  D
&=
 -Q
\\
  D
&=
  \frac{1}{\omega^2 - \Omega^2}
  Q
\qquad
=
  \frac{1}{\left(\omega-\Omega\right)\left(\omega+\Omega\right)}
\\
&=
  \frac{-1}{2\Omega}
  \left[
    \frac{1}{\omega+\Omega}
   -
    \frac{1}{\omega-\Omega}
  \right]
  Q
.
\end{align}
Now $(\omega-\Omega)^{-1}$ is a forward-like (Green's function) propagator 
 for the field, 
 but note that in my terminology, 
 it \emph{evolves} the field.
The complementary backward-like propagator
 is $(\omega+\Omega)^{-1}$.
As already described in the main text, 
 we now write ${\pDfield}={\pDfield}^++{\pDfield}^-$, 
 and split the two sides up to get
~
\begin{align}
  {\pDfield}^+
 +
  {\pDfield}^-
&=
  \frac{-1}{2\beta}
  \left[
    \frac{1}{\omega+\Omega}
   -
    \frac{1}{\omega-\Omega}
  \right]
  Q
\\
  {\pDfield}^\pm
&=
  \frac{\pm1}{2\Omega}
    \frac{1}{\omega\mp\Omega}
  Q
\\
  \left[
   \omega \mp \Omega
  \right]
  {\pDfield}^\pm
&=
 \pm
  \frac{1}{2\Omega}
    \frac{1}{\omega\mp\Omega}
  Q
\\
  \imath
  \omega {\pDfield}^\pm
&= 
 \pm
  \imath
  \Omega {\pDfield}^\pm
 \pm
  \frac{\imath}{2\Omega}
  Q
.
\end{align}
Finally, 
 we transform the frequency space $\imath \omega$ terms 
 back into normal time to give $t$ derivatives, 
 resulting in the final form
~
\begin{align}
  \partial_t
  {\pDfield}^\pm
&= 
 \pm
  \imath
  \Omega {\pDfield}^\pm
 \pm
  \frac{\imath}{2\Omega}
  Q
.
\end{align}


%
\section{The no accumulation approximation}
\label{S-RWA}

In the main text, 
 I describe the no accumulation approximation 
 in spectral terms as a RWA approximation.
However, 
 it is hard to set a clear, 
 accurate criterion for the RWA approximation to be satisfied
 in the general case, 
 since it requires knowledge of the entire propagation 
 before it can be justified.
In this appendix, 
 I take a different approach to determine
 the conditions under which the approximation will be satisfied. 
This presentation closely matches that 
 for the earlier spatial propagation version 
 by Kinsler \cite{Kinsler-2010pra-fchhg},
 but has been changed to match the temporal propagation used here.

First, 
 consider a forward evolving field
 so ${\pDfield} = {\pDfield}_0 \exp (\imath \omega t)$, 
 and therefore 
~
\begin{align}
  {\pDfield}_0^-
&=
  \frac{\omega-\Omega}{\omega+\Omega}
  {\pDfield}_0^+
\qquad
=
  \xi
  {\pDfield}_0^+
,
\end{align}
where as noted $\omega$ can be difficult to determine, 
 and may even change dynamically; 
 here we can assume it corresponds to the propagation wavevector
 that would be seen at if all the conditions holding at a chosen position
 also held everywhere else.
On this basis, 
 we can (might) even (try to) define $\omega = \omega(t)$, 
 where by analogy to the linear case we might assert that 
 $\omega^2(t) = \Omega^2 + \mathscr{Q}(t)/D(t)$, 
 so that for small $\mathscr{Q}$, 
 we have
 $\omega D \simeq \Omega ( D + \mathscr{Q} / 2 \Omega^2)$.

Let us start by assuming 
 our field is propagating and evolving forwards (only), 
 with perfectly matched ${\pDfield}^\pm$ fields; 
 so that ${\pDfield}^- = \xi {\pDfield}^+$.
 but then it happens that $\mathscr{Q}$ changes by $\delta \mathscr{Q}$
 over a small interval $\delta t$,
 likewise $\xi$ changes by $\delta \xi$.
The ${\pDfield}^\pm$ will no longer be matched, 
 and now the total field splits into two parts
 that evolve in opposite directions.
The part that continues to evolve forward has ${\pDfield}^+$ nearly unchanged, 
 but the forward evolving ${\pDfield}^-$ has changed size 
 (and is now $\propto (\xi - \delta \xi)$)
 to stay perfectly matched according to the new $\mathscr{Q}$.
The rest of the old ${\pDfield}^-$ ($\propto \delta \xi$) now propagates backwards, 
 taking with it a tiny fraction of the original ${\pDfield}^+$ 
 (and is $\propto \xi \delta \xi$).

Comparing the two backward evolving ${\pDfield}^-$ components 
 at $t$ and $t + \delta t$,
 and taking the limit $\delta t \rightarrow 0$
 enables us to estimate that the backward evolving ${\pDfield}^-$ field
 changes according to 
~
\begin{align}
  \partial_t
  {\pDfield}_{0,backward}^-
&=
  \frac{2\Omega}{\left(\omega+\Omega\right)^2}
  \left[
    \partial_t \omega
  \right]
  {\pDfield}_{0,forward}^+
.
\end{align}

Now, 
 using the small-$\mathscr{Q}$ approximation for $\omega$,
 we can write 
~
\begin{align}
  \partial_z
  {\pDfield}_{0,backward}^-
&=
  \frac{1}{\left(\omega+\Omega\right)^2}
  \left[
    \partial_t
    \mathscr{Q}
  \right]
   e^{-\imath \omega t}
.
\end{align}
where the exponential part removes any oscillations
 due to the linear part of $\mathscr{Q}$; 
 i.e. if $\mathscr{Q} = \chi {\pDfield}$ then 
~
\begin{align}
  \partial_t
  {\pDfield}_{0,backward}^-
&=
  \frac{1}{\left(\omega+\Omega\right)^2}
  \left[
    \partial_t\chi
  \right]
.
\end{align}
So here we see that backward evolving fields are only generated
 from forward evolving fields due to changes
 in the underlying conditions 
 (i.e. either material response or pulse properties), 
 but that for the reflection to be strong those changes will have to be
 significant on the order of a period, 
 or be periodic so that phase matching of the the backward wave
 could occur.

\end{document}